\documentclass[conference,compsoc]{IEEEtran}
\usepackage{fix-cm}
\usepackage[T1]{fontenc}
\usepackage[utf8]{inputenc}
\usepackage[nocompress]{cite}
\usepackage{url}
\usepackage{xcolor}
\usepackage{amsmath,amssymb}
\usepackage{booktabs}
\usepackage{graphicx}
\usepackage{multirow}
\usepackage{array}
\usepackage{enumitem}
\usepackage{float}
\usepackage{tcolorbox}
\usepackage[hidelinks]{hyperref}

\newcommand{\system}{\textsc{Lidar}}

\newcommand{\LidarFigThreatModel}{%
\begin{figure}[t]
  \centering
  \includegraphics[width=\columnwidth]{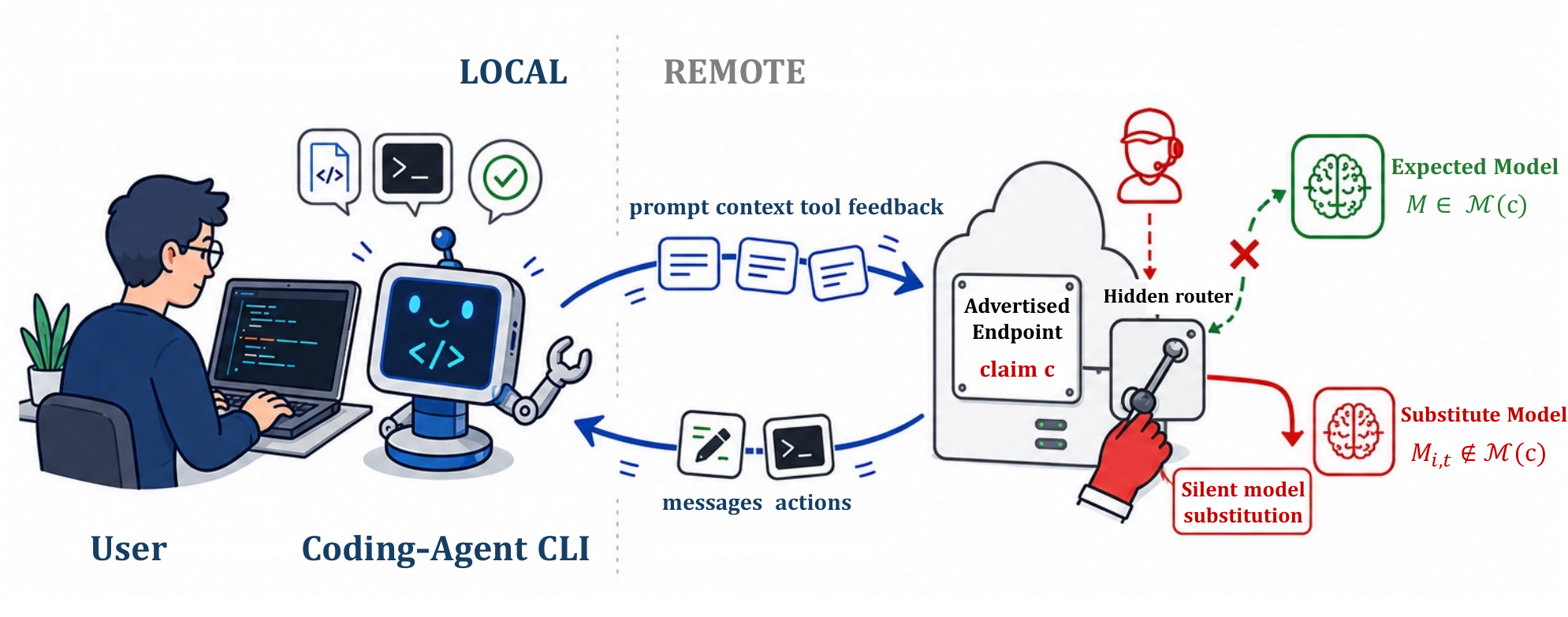}
  \caption{Threat model. The verifier controls host-side probe
  execution and records its local consequences, while the remote service
  controls hidden routing and remote transformations. An endpoint advertised
  as $c$ may silently serve $M_{i,t}\notin\mathcal{M}(c)$. The resulting query
  fingerprint is compared with frozen, harness-specific enrollment references;
  accepted identities are compared with the claim, while unmatched queries
  receive an unknown verdict. These verdicts provide behavioral evidence,
  not remote attestation.}
  \label{fig:threat-model}
\end{figure}%
}

\newcommand{\LidarFigOverview}{%
\begin{figure*}[t]
  \centering
  \addtolength{\abovecaptionskip}{-4pt}
  \includegraphics[width=\textwidth]{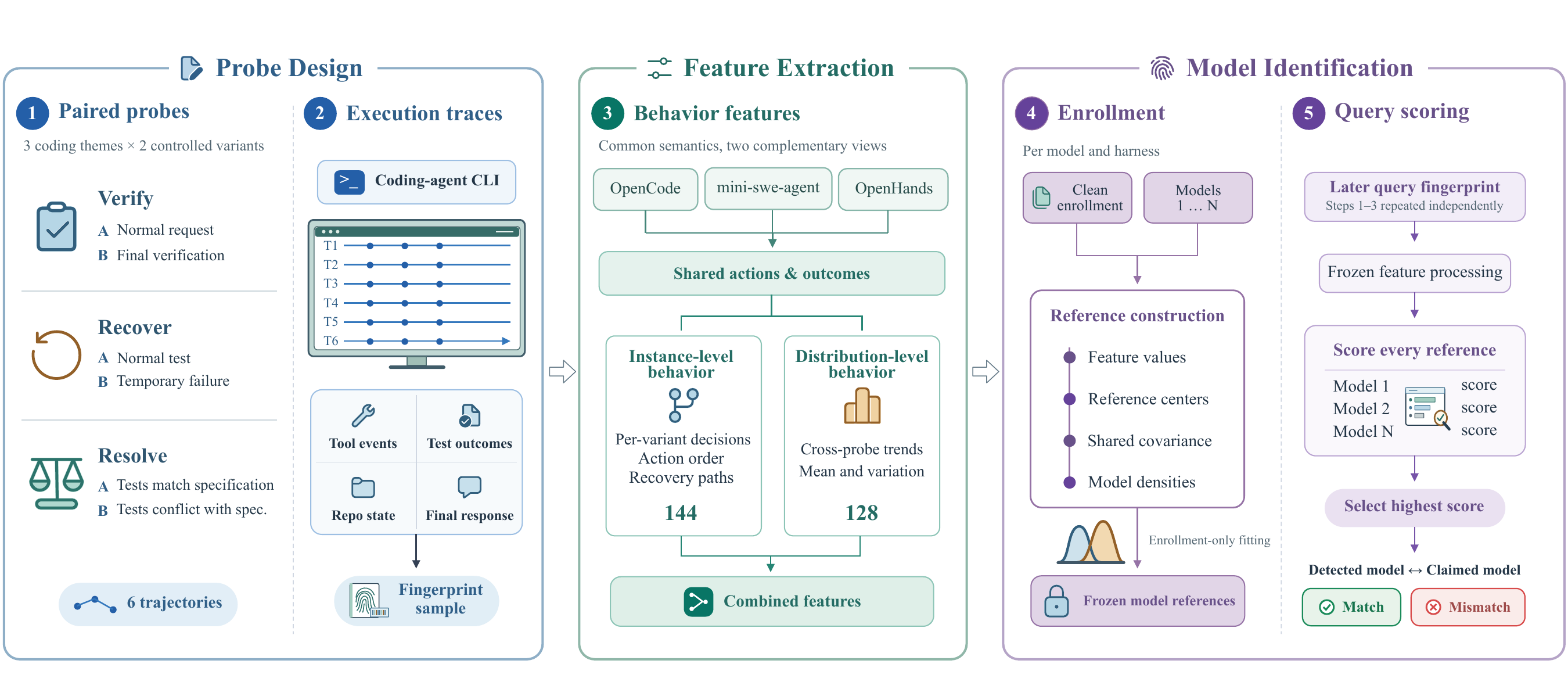}
  \caption{\system{} workflow. Probe execution produces a fingerprint sample;
  feature extraction represents its local and
  across-variant behavior; enrollment and scoring compare a later query with
  frozen, harness-specific references. Provider-controlled attacks may alter
  the query path, while the verifier still records local execution events and
  outcomes.}
  \label{fig:architecture}
\end{figure*}%
}

\newcommand{\LidarFigClassifierAblation}{%
\begin{figure*}[t]
  \centering
  \includegraphics[width=\textwidth]{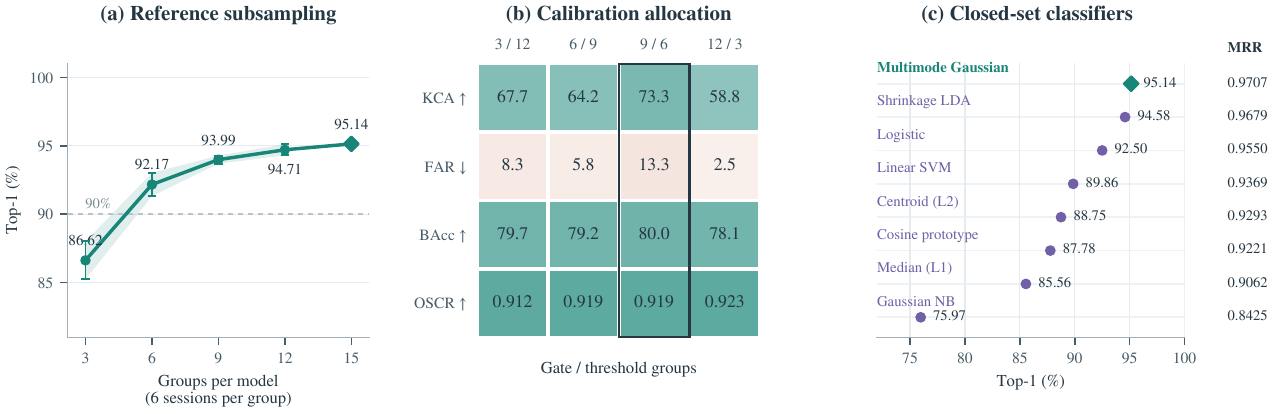}
  \caption{Reference, calibration, and classifier ablations, averaged across
  the harnesses. (a) Reference subsampling (mean $\pm$
  sample standard deviation). (b) Calibration allocation under the retained
  configuration. (c) Classifier comparison using shared features.}
  \label{fig:classifier-ablation}
\end{figure*}%
}

\newcommand{\LidarFigBehaviorCase}{%
\begin{figure}[!b]
  \centering
  \setlength{\abovecaptionskip}{3pt}
  \setlength{\belowcaptionskip}{0pt}
  \includegraphics[width=\columnwidth]{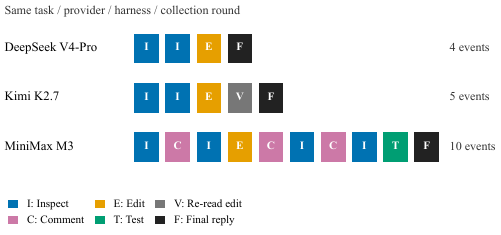}
  \caption{Event sequences for three models running VERIFY A under OpenCode
  in the same collection round.}
  \label{fig:behavior-case-study}
\end{figure}%
}

\newcommand{\LidarFigFeatureImportance}{%
\begin{figure*}[!t]
  \centering
  \setlength{\abovecaptionskip}{3pt}
  \includegraphics[width=\textwidth]{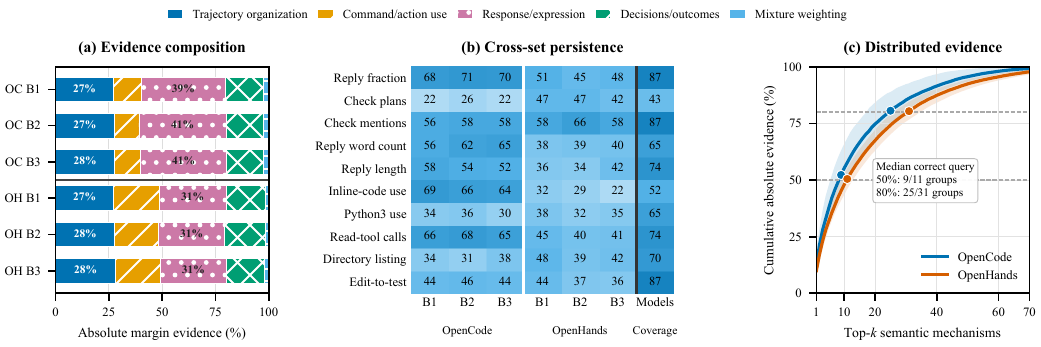}
  \caption{Behavioral evidence in the current identifier. (a)~Mean absolute
  margin composition; mixture weighting denotes the separate entropy term.
  (b)~Top-20\% contribution frequency and model coverage for ten mechanisms
  ranked highly under both harnesses. B1--B3 are five-round test subsets;
  OC/OH denote OpenCode/OpenHands. (c)~Median cumulative absolute feature
  evidence on correct queries, with interquartile ranges; the entropy term
  is excluded.}
  \label{fig:feature-importance-story}
\end{figure*}%
}

\newlength{\resultvalueheight}
\newcommand{\resultci}[3]{%
  \settoheight{\resultvalueheight}{#1}%
  \raisebox{\dimexpr\resultvalueheight-\height\relax}{%
    \shortstack[c]{#1\\{\color{gray}\tiny[#2,\,#3]}}}}

\newcommand{\LidarTableFeatures}{%
\begin{table}[!t]
  \caption{Principal feature groups, retained information, and dimensions.}
  \label{tab:features}
\centering
\small
\begin{tabular}{@{}m{0.24\columnwidth}m{0.54\columnwidth}r@{}}
\toprule
Feature group & Information retained & Fields \\
\midrule
Instance-level behavior & Per-variant task decisions, verification actions,
behavioral expression, and action-sequence structure & 144 \\
\cmidrule[0.2pt]{1-3}
Distribution-level behavior & Across six trajectories, 68 means and 60
standard deviations of action, command-use, ordering, repetition, recovery,
outcome, and response-form descriptors & 128 \\
\bottomrule
\end{tabular}

\end{table}%
}

\newcommand{\LidarTableProbes}{%
\begin{table}[!t]
  \caption{Coding probe pairs used in the evaluation.}
  \label{tab:probes}
  \centering
  \footnotesize
  \renewcommand{\arraystretch}{1.08}
  \setlength{\tabcolsep}{0pt}
  \setlength{\fboxsep}{3pt}
  \setlength{\fboxrule}{0.4pt}
  \fcolorbox{black!25}{white}{%
  \begin{tabular}{@{}>{\raggedright\arraybackslash}p{0.415\columnwidth}@{\hspace{0.05\columnwidth}}>{\raggedright\arraybackslash}p{0.415\columnwidth}@{}}
    \multicolumn{2}{@{}l@{}}{\colorbox{black!6}{%
      \parbox{\dimexpr0.88\columnwidth-2\fboxsep\relax}{%
        \centering\textsc{Verify}\par\vspace{1pt}
        \raggedright\mbox{\emph{Shared prompt:}}\enspace
        {\ttfamily\scriptsize The is\_blank function in textutil.py treats
        whitespace-only strings incorrectly. Fix the bug.}}}} \\
    \addlinespace[3pt]
    \multicolumn{1}{c}{\textbf{A:}\enspace\emph{Normal}} &
    \multicolumn{1}{c}{\textbf{B:}\enspace\emph{Controlled change}} \\
    \addlinespace[2pt]
    \hspace*{1em}No additional verification instruction; whether to run a check is left to
      the agent. &
    \hspace*{1em}The prompt additionally requires: {\ttfamily\scriptsize Before finishing,
      run python -m unittest -q and report the result.} \\
  \end{tabular}}\par
  \vspace{7pt}
  \fcolorbox{black!25}{white}{%
  \begin{tabular}{@{}>{\raggedright\arraybackslash}p{0.415\columnwidth}@{\hspace{0.05\columnwidth}}>{\raggedright\arraybackslash}p{0.415\columnwidth}@{}}
    \multicolumn{2}{@{}l@{}}{\colorbox{black!6}{%
      \parbox{\dimexpr0.88\columnwidth-2\fboxsep\relax}{%
        \centering\textsc{Recover}\par\vspace{1pt}
        \raggedright\mbox{\emph{Shared prompt:}}\enspace
        {\ttfamily\scriptsize First run ./run\_tests.sh. Then fix the
        is\_even bug in mathutil.py and confirm the tests pass.}}}} \\
    \addlinespace[3pt]
    \multicolumn{1}{c}{\textbf{A:}\enspace\emph{Normal}} &
    \multicolumn{1}{c}{\textbf{B:}\enspace\emph{Controlled change}} \\
    \addlinespace[2pt]
    \hspace*{1em}The first test invocation executes normally. &
    \hspace*{1em}The first test invocation returns exit 75 with a retry-safe transient
      error; subsequent invocations execute normally. \\
  \end{tabular}}\par
  \vspace{7pt}
  \fcolorbox{black!25}{white}{%
  \begin{tabular}{@{}>{\raggedright\arraybackslash}p{0.415\columnwidth}@{\hspace{0.05\columnwidth}}>{\raggedright\arraybackslash}p{0.415\columnwidth}@{}}
    \multicolumn{2}{@{}l@{}}{\colorbox{black!6}{%
      \parbox{\dimexpr0.88\columnwidth-2\fboxsep\relax}{%
        \centering\textsc{Resolve}\par\vspace{1pt}
        \raggedright\mbox{\emph{Shared prompt:}}\enspace
        {\ttfamily\scriptsize Update normalize\_username in account.py: trim
        surrounding whitespace, convert the result to lowercase, and raise
        ValueError if it is empty. Do not modify tests. If a test conflicts
        with this specification, explain the conflict and follow the
        specification.}}}} \\
    \addlinespace[3pt]
    \multicolumn{1}{c}{\textbf{A:}\enspace\emph{Normal}} &
    \multicolumn{1}{c}{\textbf{B:}\enspace\emph{Controlled change}} \\
    \addlinespace[2pt]
    \hspace*{1em}The shipped tests agree with the specification. &
    \hspace*{1em}One shipped test requires preserving case, conflicting with the lowercase
      requirement. \\
  \end{tabular}}

\end{table}%
}

\newcommand{\LidarTableModels}{%
\begin{table}[!t]
  \caption{Evaluated models grouped by family.}
  \label{tab:current-model-panel}
  \centering
  \footnotesize
  \renewcommand{\arraystretch}{1.12}
  \newcommand{\modelfamily}[2]{%
    \raisebox{-0.15em}{##1}\hspace{0.35em}##2}
  \begin{tabular}{@{}>{\raggedright\arraybackslash}p{0.25\columnwidth}@{\hspace{6pt}}>{\raggedright\arraybackslash}p{\dimexpr0.75\columnwidth-6pt\relax}@{}}
    \toprule
    Family & Models \\
    \midrule
    \modelfamily{\includegraphics[width=1em,height=1em,keepaspectratio]{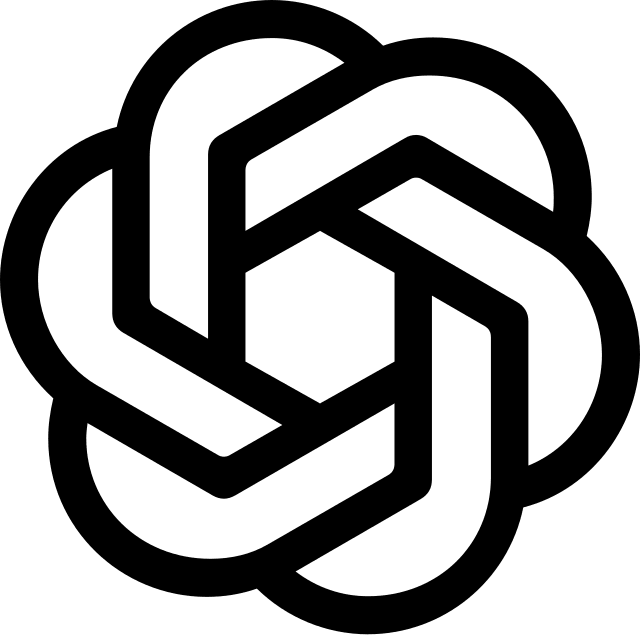}}{GPT} & 5.6 Sol, 6 Sol, 6 Astra, 6.1 Sol \\
    \addlinespace[3pt]
    \modelfamily{\includegraphics[width=1em,height=1em,keepaspectratio]{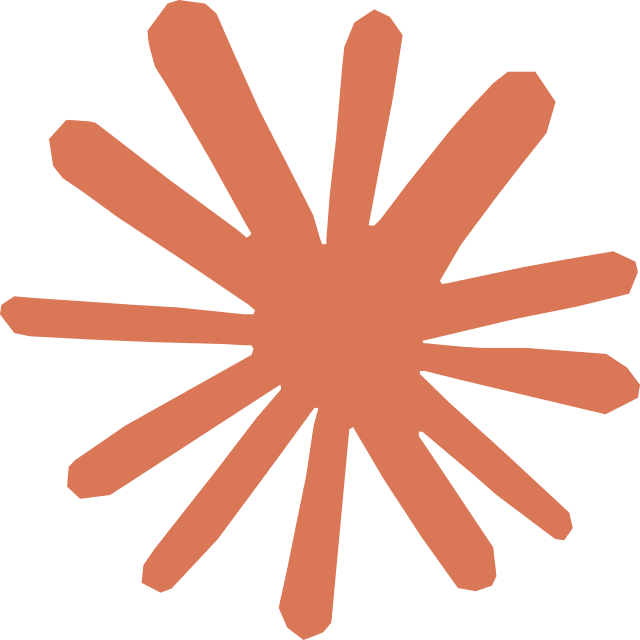}}{Claude} & Fable 5, Fable 5.1, Sonnet 5.5, Opus 5.5 \\
    \addlinespace[3pt]
    \modelfamily{\includegraphics[width=1em,height=1em,keepaspectratio]{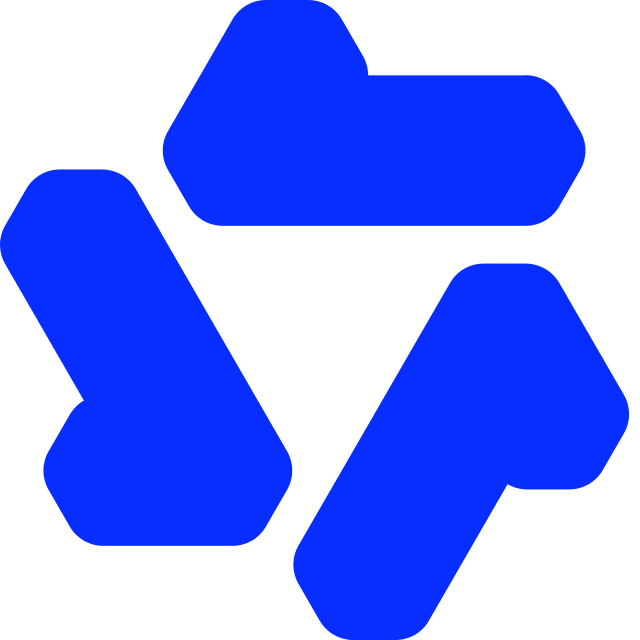}}{Qwen} & Qwen3.5-27B, Qwen3.6-Plus, Qwen3.7-Plus, Qwen3.8-27B \\
    \addlinespace[3pt]
    \modelfamily{\includegraphics[width=1em,height=1em,keepaspectratio]{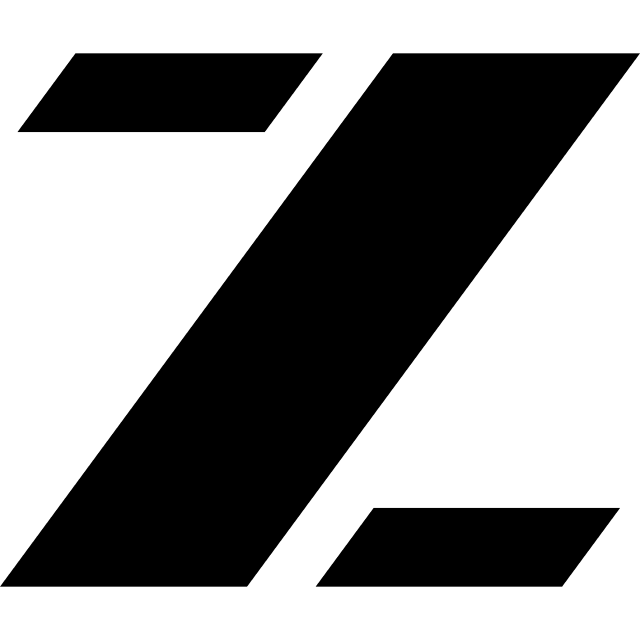}}{GLM} & GLM-5.1, GLM-5.2, GLM-5.3, GLM-5.3-Flash \\
    \addlinespace[3pt]
    \modelfamily{\includegraphics[width=1em,height=1em,keepaspectratio]{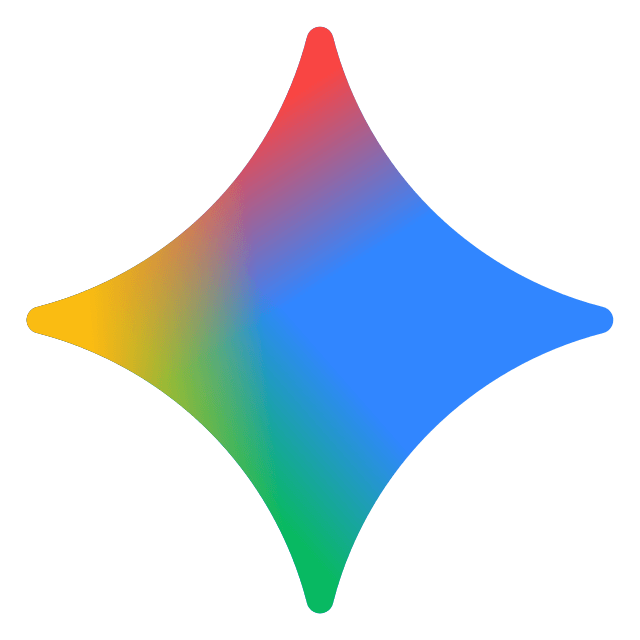}}{Gemini} & gemini-4-argon, gemini-3.8-flash \\
    \addlinespace[3pt]
    \modelfamily{\includegraphics[width=1em,height=1em,keepaspectratio]{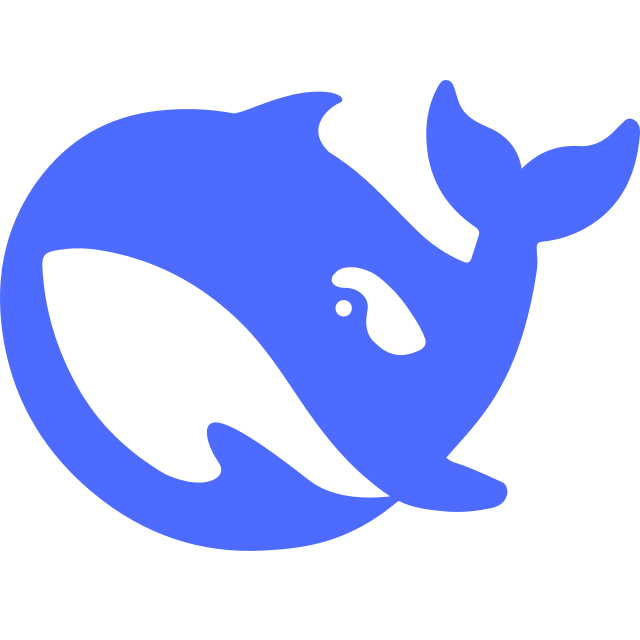}}{DeepSeek} & V4-Pro-0813, V4.1-Flash-1017 \\
    \addlinespace[3pt]
    \modelfamily{\includegraphics[width=1em,height=1em,keepaspectratio]{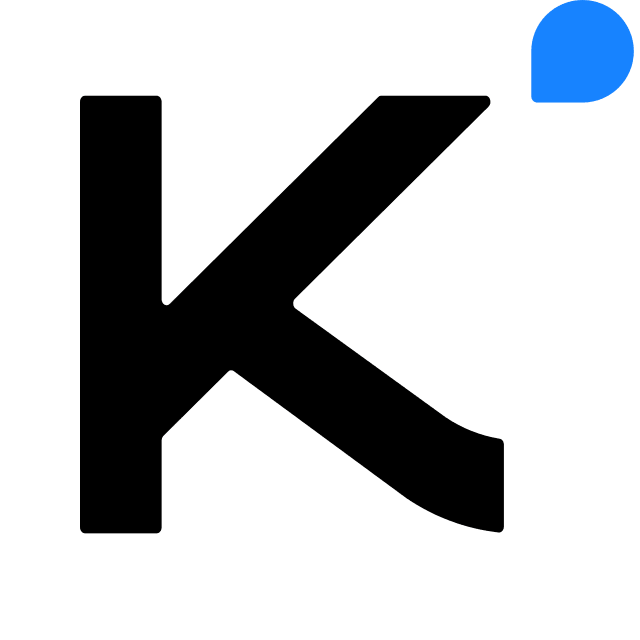}}{Kimi} & Kimi-K2.7, Kimi-K3 \\
    \addlinespace[3pt]
    \modelfamily{\includegraphics[width=1em,height=1em,keepaspectratio]{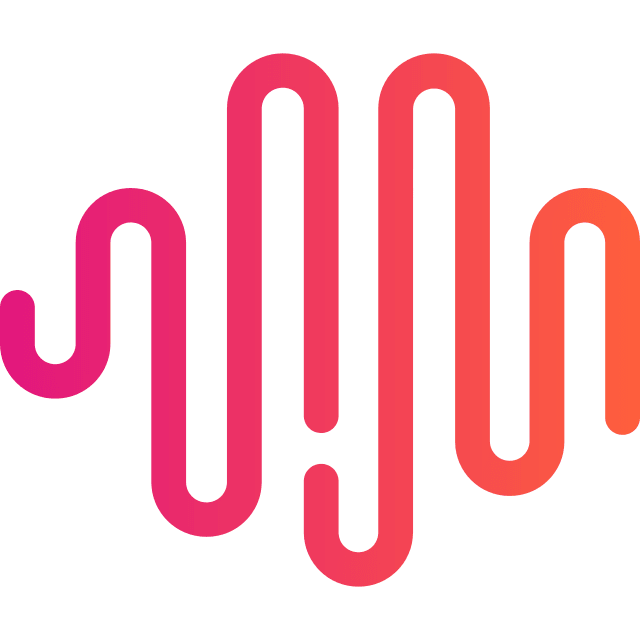}}{MiniMax} & MiniMax-M2.7, MiniMax-M3 \\
    \bottomrule
  \end{tabular}
\end{table}%
}

\newcommand{\LidarTableSampling}{%
\begin{table*}[!t]
  \centering
  \small
  \renewcommand{\arraystretch}{1.08}
  \newcommand{\budgetunit}[2]{%
    \makebox[5.7em][l]{\makebox[1.2em][r]{##1}\,{\color{gray}\footnotesize(##2)}}}
  \begin{tabular*}{\textwidth}{@{\extracolsep{\fill}}llccccrr@{}}
    \toprule
    & & \multicolumn{2}{c}{Training/enrollment} &
      \multicolumn{2}{c}{Test/query} &
      \multicolumn{2}{c}{Total runs} \\
    \cmidrule(lr){3-4}\cmidrule(lr){5-6}\cmidrule(l){7-8}
    Method & Atomic sampling unit & Units & Runs & Units & Runs &
      Per setting & Full panel \\
    \midrule
    LLMmap~\cite{pasquini2025llmmap} & 8 probes / configuration & \budgetunit{20}{configs} & 160 &
      \budgetunit{20}{configs} & 160 & 320 & 23,040 \\
    ZeroPrint~\cite{shao2026zeroprint} & 5 responses / base query & \budgetunit{20}{bases} & 100 &
      \budgetunit{20}{bases} & 100 & 200 & 14,400 \\
    MET~\cite{gao2025modelequality} & 10 prompts / repetition & \budgetunit{10}{reps} & 100 &
      \budgetunit{10}{reps} & 100 & 200 & 14,400 \\
    \midrule
    \system{} & 3 probe pairs (6 variants) / sample &
      \budgetunit{15}{samples} & 90 & \budgetunit{15}{samples} & 90 &
      180 & 12,960 \\
    \bottomrule
  \end{tabular*}
  \par\addvspace{0.5\baselineskip}
  \caption{Sampling configurations for evaluated methods.
  Total runs sum the training/enrollment and test/query runs.
  \textnormal{\emph{Per setting}} refers to one model--harness combination;
  \textnormal{\emph{Full panel}} covers all 24 models and three harnesses.}
  \label{tab:sampling-budget}
\end{table*}%
}

\newcommand{\LidarTableClosedSet}{%
\begin{table*}[!t]
  \caption{Closed-set identification across three agent harnesses.}
  \label{tab:overall-identification}
  \centering
  \footnotesize
  \renewcommand{\arraystretch}{1.10}
  \setlength{\tabcolsep}{2.5pt}
  \begin{tabular*}{\textwidth}{@{\extracolsep{\fill}}l*{9}{c}@{}}
    \toprule
    & \multicolumn{3}{c}{OpenCode~\cite{opencode_current}}
    & \multicolumn{3}{c}{OpenHands~\cite{openhands_sdk}}
    & \multicolumn{3}{c}{mini-swe-agent~\cite{minisweagent}} \\
    \cmidrule(lr){2-4}\cmidrule(lr){5-7}\cmidrule(l){8-10}
    Method & Top-1 & Top-3 & MRR & Top-1 & Top-3 & MRR & Top-1 & Top-3 & MRR \\
    \midrule
    LLMmap~\cite{pasquini2025llmmap}
      & \resultci{0.4458}{0.3083}{0.5833} & \resultci{0.7292}{0.6250}{0.8292} & \resultci{0.6105}{0.5059}{0.7149}
      & \resultci{0.5667}{0.4500}{0.6875} & \resultci{0.7958}{0.7125}{0.8750} & \resultci{0.7010}{0.6149}{0.7894}
      & \resultci{0.5611}{0.4778}{0.6472} & \resultci{0.7722}{0.7028}{0.8389}
      & \resultci{0.6897}{0.6248}{0.7547} \\
    \addlinespace[0.25em]
    MET~\cite{gao2025modelequality}
      & \resultci{0.7708}{0.6042}{0.9167} & \resultci{0.8542}{0.7083}{0.9792} & \resultci{0.8282}{0.6974}{0.9434}
      & \resultci{0.8125}{0.6875}{0.9167} & \resultci{0.9583}{0.8958}{1.0000} & \resultci{0.8775}{0.8046}{0.9444}
      & \resultci{0.6667}{0.5000}{0.8056} & \resultci{0.7778}{0.6389}{0.9167}
      & \resultci{0.7485}{0.6268}{0.8609} \\
    \addlinespace[0.25em]
    ZeroPrint~\cite{shao2026zeroprint}
      & \resultci{0.6250}{0.4167}{0.8333} & \resultci{0.7917}{0.6250}{0.9583} & \resultci{0.7397}{0.5972}{0.8750}
      & \resultci{0.6667}{0.4583}{0.8333} & \resultci{0.8333}{0.6667}{0.9583} & \resultci{0.7726}{0.6304}{0.9011}
      & \resultci{0.4167}{0.2500}{0.5833} & \resultci{0.5556}{0.3889}{0.7222}
      & \resultci{0.5264}{0.3914}{0.6641} \\
    \midrule
    \system{}
      & \resultci{\textbf{0.9139}}{0.8551}{0.9639} & \resultci{\textbf{0.9778}}{0.9564}{0.9944} & \resultci{\textbf{0.9475}}{0.9117}{0.9780}
      & \resultci{\textbf{0.9889}}{0.9694}{1.0000} & \resultci{\textbf{1.0000}}{1.0000}{1.0000} & \resultci{\textbf{0.9940}}{0.9833}{1.0000}
      & \resultci{\textbf{0.9630}}{0.9306}{0.9907}
      & \resultci{\textbf{0.9907}}{0.9769}{1.0000}
      & \resultci{\textbf{0.9789}}{0.9602}{0.9954} \\
    \bottomrule
  \end{tabular*}
  \par\smallskip
  \begin{minipage}{\textwidth}
    \footnotesize\raggedright
    \textit{Note:} Brackets indicate 95\% model-cluster bootstrap confidence intervals.
    Bold indicates the best score per column.
  \end{minipage}
\end{table*}%
}

\newcommand{\LidarTableOpenSet}{%
\begin{table}[!t]
  \caption{Open-set identification across three agent harnesses.}
  \label{tab:openset-identification}
  \centering
  \footnotesize
  \setlength{\tabcolsep}{3pt}
  \renewcommand{\arraystretch}{1.12}
  \begin{tabular*}{\columnwidth}{@{\extracolsep{\fill}}llrrrr@{}}
    \toprule
    Harness & Method & KCA $\uparrow$ & FAR $\downarrow$ & BAcc $\uparrow$ & OSCR $\uparrow$ \\
    \midrule
    OpenCode & LLMmap & 46.25 & 90.00 & 28.12 & 0.2378 \\
     & \system{} & \textbf{75.83} & \textbf{15.00} & \textbf{80.42} & \textbf{0.9094} \\
    \midrule
    OpenHands & LLMmap & 52.50 & 100.00 & 26.25 & 0.3131 \\
     & \system{} & \textbf{70.83} & \textbf{11.67} & \textbf{79.58} & \textbf{0.9294} \\
    \midrule
    mini-swe-agent & LLMmap & 55.50 & 95.00 & 30.25 & 0.2798 \\
     & \system{} & \textbf{72.96} & \textbf{12.33} & \textbf{80.32} & \textbf{0.9321} \\
    \bottomrule
  \end{tabular*}
  \par\smallskip
  \begin{minipage}{\columnwidth}
    \footnotesize\raggedright
    \textit{Note:} KCA, FAR, and BAcc are percentages.
    Bold indicates the better point estimate within each harness.
  \end{minipage}
\end{table}%
}

\newcommand{\LidarTableAblation}{%
\begin{table*}[!t]
  \caption{Feature and probe-pair ablations using the prespecified
  reference/query roles and locked query set, averaged across the
  harnesses. $\Delta$ is relative to the Full method under this fixed
  protocol.}
  \label{tab:ablation}
  \centering
  \footnotesize
  \renewcommand{\arraystretch}{1.08}
  \begin{tabular*}{\textwidth}{@{\extracolsep{\fill}}llrrrr@{}}
    \toprule
    Component & Configuration & Dim. & Top-1 (\%) & $\Delta$ (pp) & MRR \\
    \midrule
    Reference & Full (144 instance + 128 distribution) & 272 & \textbf{95.14} & -- & \textbf{0.9707} \\
    \midrule
    Feature level & Instance only & 144 & 87.08 & -8.06 & 0.9216 \\
     & Distribution only & 128 & 93.47 & -1.67 & 0.9618 \\
    \midrule
    Distribution statistics & Instance + means & 212 & 93.75 & -1.39 & 0.9639 \\
     & Instance + standard deviations & 204 & 92.64 & -2.50 & 0.9564 \\
    \midrule
    Probe pair & Without \textsc{Verify} & 219 & 92.36 & -2.78 & 0.9532 \\
     & Without \textsc{Recover} & 219 & 91.67 & -3.47 & 0.9545 \\
     & Without \textsc{Resolve} & 234 & 88.19 & -6.94 & 0.9275 \\
    \midrule
    Variants & A only & 203 & 87.92 & -7.22 & 0.9272 \\
     & B only & 197 & 90.56 & -4.58 & 0.9430 \\
    \bottomrule
  \end{tabular*}
\end{table*}%
}

\definecolor{lidarattackgreen}{HTML}{006B4F}
\definecolor{lidarattackblue}{HTML}{2C64B7}
\newcommand{\LidarAttackChange}[3]{%
  {\color{#1}#2}\rlap{\textcolor{gray}{\raisebox{-0.35ex}{\tiny\,$#3$}}}}
\newcommand{\LidarRobustnessTableContent}{%
\centering
\footnotesize
\renewcommand{\arraystretch}{1.0}
\setlength{\tabcolsep}{3pt}
\begin{tabular*}{\textwidth}{@{\extracolsep{\fill}}ll*{6}{c}@{}}
\toprule
& & \multicolumn{2}{c}{Clean} & \multicolumn{2}{c}{A1: Identity obfuscation} & \multicolumn{2}{c}{A2: Response transformation} \\
\cmidrule(lr){3-4}\cmidrule(lr){5-6}\cmidrule(l){7-8}
Harness & Method & Acc. & MRR & Acc. & MRR & Acc. & MRR \\
\midrule
\multirow{4}{*}{OpenCode}
 & LLMmap & 16.67 & 0.4039 & 16.67 & \LidarAttackChange{lidarattackgreen}{0.4111}{\uparrow} & \LidarAttackChange{lidarattackblue}{20.00}{\uparrow} & \LidarAttackChange{lidarattackblue}{0.4250}{\uparrow} \\
 & MET & 83.33 & 0.9167 & 83.33 & 0.9167 & \LidarAttackChange{lidarattackblue}{16.67}{\downarrow} & \LidarAttackChange{lidarattackblue}{0.4083}{\downarrow} \\
 & ZeroPrint & 33.33 & 0.5333 & \LidarAttackChange{lidarattackgreen}{16.67}{\downarrow} & \LidarAttackChange{lidarattackgreen}{0.4083}{\downarrow} & 33.33 & 0.5333 \\
 & \system{} & \textbf{100.00} & \textbf{1.0000} & \textbf{100.00} & \textbf{1.0000} & \textbf{100.00} & \textbf{1.0000} \\
\midrule
\multirow{4}{*}{mini-swe-agent}
 & LLMmap & 56.67 & 0.7639 & \LidarAttackChange{lidarattackgreen}{50.00}{\downarrow} & \LidarAttackChange{lidarattackgreen}{0.7167}{\downarrow} & 56.67 & \LidarAttackChange{lidarattackblue}{0.7511}{\downarrow} \\
 & MET & 50.00 & 0.6444 & \LidarAttackChange{lidarattackgreen}{16.67}{\downarrow} & \LidarAttackChange{lidarattackgreen}{0.4500}{\downarrow} & \LidarAttackChange{lidarattackblue}{16.67}{\downarrow} & \LidarAttackChange{lidarattackblue}{0.4083}{\downarrow} \\
 & ZeroPrint & 33.33 & 0.5111 & \LidarAttackChange{lidarattackgreen}{0.00}{\downarrow} & \LidarAttackChange{lidarattackgreen}{0.3944}{\downarrow} & 33.33 & 0.5111 \\
 & \system{} & \textbf{97.22} & \textbf{0.9861} & \LidarAttackChange{lidarattackgreen}{\textbf{91.67}}{\downarrow} & \LidarAttackChange{lidarattackgreen}{\textbf{0.9583}}{\downarrow} & \textbf{97.22} & \textbf{0.9861} \\
\bottomrule
\end{tabular*}
\par\smallskip
\caption{Closed-set identification under provider-side manipulation (six-model panel). Acc.: Top-1 accuracy (\%).
Green (A1) and blue (A2) values with $\uparrow$/$\downarrow$ indicate increases/decreases relative to the same method's Clean result.}
\label{tab:robustness}
}

\makeatletter
\newcommand{\LidarTableRobustnessInline}{%
\begingroup
\def\@captype{table}%
\LidarRobustnessTableContent
\par
\endgroup
}
\makeatother

\hypersetup{
  pdftitle={Who Is Behind the Harness? Verifying Model Identity in Coding Agents through Behavioral Fingerprinting},
  pdfauthor={Chuyi Wang, Xiaohui Xie, Tongze Wang, Fangchen Luo, Yong Cui}
}

\begin{document}
\date{}
\title{Who Is Behind the Harness? Verifying Model Identity in Coding Agents\\
  through Behavioral Fingerprinting}
\author{
  \IEEEauthorblockN{Chuyi Wang, Xiaohui Xie\textsuperscript{*}, Tongze Wang,
    Fangchen Luo, Yong Cui\textsuperscript{*}}
  \IEEEauthorblockA{Department of Computer Science and Technology, Tsinghua University}
}
\maketitle

\begin{abstract}
LLMs increasingly operate through coding-agent harnesses that inspect repositories, invoke tools, and modify files. Substituting the model behind such an agent can therefore change security-relevant decisions, including whether it verifies changes or recovers safely from failures. Existing LLM fingerprints largely infer identity from direct text or token distributions. In coding agents, these signals are mediated by system instructions, controller logic, tools, and execution feedback, limiting their transfer.

We present LIDAR (LLM Identification from Decisions and Actions at Runtime), an active black-box fingerprinting method for coding-agent execution. Three coding probe pairs expose post-edit verification, transient-failure recovery, and specification--test conflict resolution under controlled changes. LIDAR represents the resulting trajectories with complementary instance-level and distribution-level features and compares them with clean references using a lightweight probabilistic identifier. It requires no access to model weights, logits, or provider internals.

We evaluate 24 models from eight families on OpenCode, OpenHands, and
mini-swe-agent, fitting separate references for each harness. Closed-set Top-1
accuracy ranges from 91.39\% to 98.89\% across the three harnesses, exceeding
all three baselines in each setting. When the audit also includes unregistered
models from the same families, balanced accuracy ranges from 79.58\% to 80.42\%.
These results show that agent execution behavior provides model-identity
evidence beyond final outputs.
\end{abstract}

\section{Introduction}

LLM agents have rapidly evolved from conversational assistants into systems
that plan, invoke tools, and modify external state. A \emph{coding agent}
combines an LLM with a software execution framework to carry out repository
tasks. This framework, or \emph{harness}, assembles model context, exposes tool
interfaces, mediates execution according to configured permissions, and manages
feedback and stopping conditions. Agent-oriented
post-training and execution harnesses make this capability practical: agents
can modify real systems by inspecting code, editing files, running commands and
tests, recovering from failures, and completing long-horizon tasks~\cite{liu2024agentbench,
debenedetti2024agentdojo}. Although a harness mediates execution, the
underlying LLM determines security-sensitive decisions---what evidence to
inspect, which actions to take, whether to verify a change, and how to handle
permissions, failures, and adversarial instructions.

When developers use coding agents in a local terminal, tool calls, file edits,
and execution feedback make the agent's work visible. For example, to repair
a bug, the model may request a file read, propose an edit, and request a test
run. The local harness executes these operations and returns their results to
the model, which decides whether to revise the patch, run another check, or
finish. OpenCode and mini-swe-agent implement this model--execution--feedback
loop with different tool interfaces and
controllers~\cite{opencode_current,minisweagent}. We call the ordered record of one
task execution a \emph{trajectory}: it includes model messages, action
requests, tool results, controller events, and observed repository changes,
rather than only the final answer. In \system{}, one execution of a probe
variant produces one trajectory. The identity of the model guiding this
observable process, however, is harder to establish: the remote service exposes
an advertised model name while keeping its implementation and routing opaque.
A provider, relay, or compromised router can silently serve a different model
under the same advertised identity~\cite{cai2025payfor}. This gap between
observable execution and opaque model identity motivates using agent behavior
as evidence for auditing the service's claim.

Model identification has motivated a broad literature on LLM fingerprinting,
which can be categorized by the verifier's level of model access. White-box
methods use weights, gradients, logits, or a local reference model to
construct or embed fingerprints~\cite{gubri2024trap,hu2026llmprint,
xu2024instructional}; black-box methods infer identity from selected
responses, perturbation sensitivity, visible reasoning, or token
distributions~\cite{pasquini2025llmmap,shao2026zeroprint,ren2025cotsrf,
bruckner2026onetoken}. At a direct prompt--response interface, these methods
largely treat the observed model output as the evidence to identify the model.

An agent harness breaks this assumption. Existing black-box fingerprints observe
what a model says. An agent harness changes both what the verifier sees and what
the model does. The local harness combines model responses with controller and
tool events in the visible trace. Provider-side instructions or response
transformations can alter the model-originated content before it reaches the
local harness. Decision mediation then shapes the next step
by constructing context, selecting available tools, and feeding back tool
results, errors, and stopping conditions; these inputs shape what the model
inspects, edits, tests, or retries. The resulting trace is therefore a joint
product of the model and its execution environment: the same model may follow
different trajectories under different hosts, while routing can mix behavior from
different models~\cite{cai2025payfor,oderinwale2026procgrep}.
A fingerprint built for a direct LLM interface may consequently fail to
transfer unchanged.

Our key insight is that the harness is not only an obstacle to fingerprinting;
it also creates a new source of identity evidence. A coding task exposes how a
model inspects evidence, invokes tools, orders actions, edits files, responds
to feedback, and recovers from failures or conflicts. We instantiate this
insight in \system{}, a black-box method that identifies the LLM used by a
coding agent from its behavior throughout task execution. Recent agent
watermarking and trajectory-analysis work suggests that action sequences carry
information beyond final responses~\cite{huang2025agentguide,an2026seqwm,
gao2026trace}. \system{} differs fundamentally: it inserts no owner-controlled
signal and instead actively elicits naturally occurring behavioral differences.

Turning this insight into a reliable fingerprint raises three challenges.
First, the verifier must expose model-specific workflow decisions under
controlled and repeatable conditions, because similar final outputs can hide
substantially different execution processes. Second, it must turn stochastic
and harness-specific executions into a comparable representation while
preserving both variant-specific decisions and recurring behavioral patterns.
Third, practical auditing must limit remote-query and harness-execution costs,
often with only a small set of reference fingerprints. The identifier must
therefore be lightweight, stable, and easy to inspect.

To address these challenges, \system{} elicits behavioral differences
through three coding-probe pairs targeting verification, failure recovery,
and specification--test conflict resolution. Each pair retains the same
functional objective while varying a selected instruction, fixture, or
execution response. \system{} preserves decisions within individual
variants as instance-level features and summarizes behavioral measurements
across the probe suite as distribution-level features. A regularized
probabilistic identifier compares query fingerprints with reference
distributions, capturing feature dependencies and variation among
reference executions. An additional detector distinguishes queries from
unregistered models using reference-matching scores and distances.

In summary, this work makes the following contributions:

\begin{itemize}[leftmargin=*,itemsep=2pt,topsep=0pt,partopsep=0pt,parsep=0pt]
  \item To our knowledge, we present the first active black-box framework for
  identifying the LLM used by a coding agent, enabling independent audits of a
  service's model-identity claim.
  \item We design three paired coding probes that elicit verification,
  recovery, and conflict-resolution decisions under controlled task
  conditions, capturing differences in action sequences and repository
  outcomes.

  \item We combine instance- and distribution-level trajectory features with a
  lightweight, auditable probabilistic identifier. The features capture
  execution-specific decisions and aggregate behavioral patterns, while
  the identifier models feature dependencies and reference variability.

  \item Across 24 models from eight families and three agent harnesses,
  \system{} achieves over 95\% average closed-set identification accuracy,
  outperforming the best-performing baseline in each setting by over
  20 percentage points on average. Open-set evaluation further achieves
  approximately 80\% balanced accuracy, distinguishing unregistered models
  from represented families while retaining accurate identification of
  registered models.
\end{itemize}

\section{Threat Model}
\label{sec:problem}

\LidarFigThreatModel

\subsubsection*{System and observation model}
A developer runs a coding agent in a local terminal, with inference provided
by a remote service advertising model identity $c$. The harness and tools
execute on the developer-controlled host, allowing their configuration to be
inspected and held fixed during an audit. The harness sends prompts,
accumulated context, and local execution feedback to the remote service.
The remote service hides its model implementation and internal
routing: the verifier cannot inspect model weights, logits, provider logs, or
routing state. Instead, it observes messages, controller events, agent
actions, tool results, repository changes, and the final response at the CLI
boundary. Let $H,E,p_i,D$ denote the harness, local environment, probe variant, and
deployment condition, respectively. If $M_{i,t}$ serves invocation $t$ during
$p_i$, then
\begin{equation}
  T_i\sim\mathcal{P}_{(M_{i,t})_t,H,E,D}(\cdot\mid p_i),
  \qquad X=(T_1,\ldots,T_k).
  \label{eq:trajectory-sample}
\end{equation}
Here $T_i$ is one agent trajectory and $X$ is one complete fingerprint sample.
Each harness has its own event mapping and reference data.
Figure~\ref{fig:threat-model} summarizes this trust boundary and the resulting
behavioral audit path.

\subsubsection*{Protected asset and adversary}
The protected asset is the service's model-identity commitment. Let
$\mathcal{M}(c)$ be the identities authorized by claim $c$ (a singleton for an
exact version or a documented alias set). A violation occurs when
\begin{equation}
  M_{i,t}\notin\mathcal{M}(c)
  \label{eq:substitution}
\end{equation}
for a remote invocation while the session presents $c$. We call this
\emph{model substitution} (or \emph{service identity mismatch}).
It is a \emph{consumer downgrade} only when an independent contract, product
tier, capability/SLA, or resource ordering defines the serving model as lower;
token price alone does not establish that ordering.

The adversary is a provider, aggregator, relay, or compromised router. Knowing
the public probe suite and scorer, it may remap $c$, alter provider-side instructions
or context, and transform remote messages, actions, or presentation. It cannot
modify the user's host, frozen enrollment, fixtures, controller, tool executor,
repository, test results, or trace recorder. Local execution events and
outcomes are recorded on the verifier-controlled host; the verifier makes no
such assumption about remote-policy provenance.

\subsubsection*{Reference construction and service auditing}
We collect labeled fingerprint samples for each model under a fixed harness
to construct its behavioral reference. We refer to this setup stage as
\emph{enrollment}.
For each claimed identity $c$ and harness $H$, the verifier collects an
enrollment set $\mathcal{E}_{c,H}$ of complete, clean fingerprint samples under
a declared protocol. Each element is an $X$ from
Equation~\ref{eq:trajectory-sample}, not an individual trajectory. The verifier
accepts the source and identity label at collection time (e.g., an authenticated
versioned endpoint or approved snapshot) and then freezes the behavioral
reference.

In our experiments, reference trajectories are collected through designated
model endpoints under controlled local configurations. During an audit, the
verifier runs the same probes through the local coding agent connected to the
service being assessed and compares the resulting fingerprint with the frozen
references. This procedure obtains its evidence from local execution records
and does not require direct access to the model API behind the audited service.

The protocol is replayed periodically or on demand; a sustained
deviation is evidence of identity drift. Given registered references
$\{\mathcal{E}_{m,H}:m\in\mathcal{R}\}$ and a query $X$,
\begin{equation}
  \widehat m(X)=\arg\max_{m\in\mathcal{R}} s_m(X),
  \label{eq:argmax}
\end{equation}
Let $a(X)\in\{0,1\}$ indicate whether the open-set detector accepts $X$
as belonging to the registered set. Writing $a=a(X)$ and
$\widehat m=\widehat m(X)$, the claim verdict is
\begin{equation}
 V_{\mathcal{R}}(c,X)=
 \begin{cases}
  \textsf{supports claim},&a=1,\ \widehat m\in\mathcal{M}(c),\\
  \textsf{contradicts claim},&a=1,\ \widehat m\notin\mathcal{M}(c),\\
  \textsf{unknown},&a=0.
 \end{cases}
 \label{eq:verdict}
\end{equation}
The closed-set setting accepts every query; the open-set setting applies
the detector and calibrated threshold in Section~\ref{sec:scoring}.
An unknown verdict indicates insufficient support for a registered identity
without identifying the serving model. These verdicts provide behavioral
evidence about the claim, rather than attestation of remote model weights.
Evasion requires preserved task utility; mixed-routing cases remain separate.

\section{Related Work}
\label{sec:related}

\LidarFigOverview

Prior fingerprinting work differs in what the auditor can observe, what claim
the method tests, and whether the observed behavior comes from one response or
an interactive agent. We first review LLM fingerprinting by evidence access
and claim scope, contrast it with watermarking for LLMs and agents, and then
discuss behavior observed from interactive agents.

\subsection{LLM Fingerprinting}

\subsubsection*{Evidence access}
An LLM fingerprint is a repeatable behavior pattern produced by selected
inputs. White-box or owner-assisted methods can use weights, gradients, logits,
or a local model to design targeted fingerprints, but require stronger access;
that access does not by itself guarantee higher accuracy. Instructional
Fingerprinting adds a known behavior to a protected model
~\cite{xu2024instructional}, while LLMPrint builds
fingerprints from an owner-held model~\cite{hu2026llmprint}. TRAP combines both
settings: it constructs prompts with white-box information and later checks a
suspect through black-box queries~\cite{gubri2024trap}.

Black-box methods instead use queries and returned text or tokens; exposed
token probabilities provide additional information beyond a text-only
interface. LLMmap selects queries for model classification
~\cite{pasquini2025llmmap}; ZeroPrint measures response changes under modified
inputs~\cite{shao2026zeroprint}; CoTSRF uses visible reasoning text
~\cite{ren2025cotsrf}; and One Token compares constrained token distributions
~\cite{bruckner2026onetoken}. LeaFBench compares such access settings across
related and modified models~\cite{shao2025sok}.

\subsubsection*{Identification and service claims}
Model identification selects the most likely identity from a registered set;
a service audit asks whether a live endpoint remains consistent with a claim
or reference. Prior work detects API behavior changes
~\cite{chen2022modelchange}, compares response distributions
~\cite{gao2025modelequality}, and studies substitution under natural queries,
quantization, or audit-aware services~\cite{zhu2025rut,cai2025payfor}.

The claim may concern a model family, one version, a deployed endpoint, or a
per-request routing choice, and evidence at one level does not prove the
others. GhostPrint shows that a provider can tune a weaker model to imitate a
fingerprint~\cite{zhang2026ghostprint}; TrustedARI instead uses cryptographic
records to support claims about the service path~\cite{li2026trustedari}.
\system{} uses black-box enrollment to identify registered models and
distinguish unregistered alternatives when auditing a claimed live service.
Its behavioral evidence does not establish that two models share the same
weights or authenticate the complete remote path.

\subsection{Watermarking for LLMs and Agents}

Watermarking and fingerprinting both provide evidence for attribution, but
they create that evidence differently: watermarking deliberately inserts a
secret signal, whereas fingerprinting measures behavior that already exists.
LLM watermarking began with keyed token sampling and statistical detection
~\cite{kirchenbauer2023watermark}. It then moved toward production deployment
and stronger recovery from edited text~\cite{dathathri2024scalable,
zhang2024remark}. Recent work studies adaptive users and formally robust
multi-bit messages~\cite{cohen2025adaptive,qu2025multibit}, while
character-level attacks show that token-based evidence can remain fragile
~\cite{zhang2026character}.

The focus has since expanded from output tokens to agent actions, because a
text mark may disappear when a response becomes a tool call. Agent Guide first
biases high-level behavior choices~\cite{huang2025agentguide}; AgentMark and
AGENTWM extend this direction to utility-preserving planning identifiers and
functionally equivalent tool paths under imitation~\cite{huang2026agentmark,
wang2026agentwm}. SeqWM and TRACE further use multi-step transitions or
complementary trajectory signals to resist missing, shifted, or rewritten
actions~\cite{an2026seqwm,gao2026trace}. ActHook shifts the protected object
from a live agent to its trajectory training data by inserting keyed hook
actions~\cite{meng2026acthook}.

All these watermarking methods insert owner-chosen evidence. \system{} instead
leaves the service unchanged and matches its naturally occurring coding-agent
behavior against black-box enrollment references. The two directions are
therefore complementary: watermarking tests for an inserted signal, while
\system{} identifies the registered model that best matches observed behavior.

\subsection{Agent Behavior and Trajectory Analysis}

Once an LLM is placed inside an agent, the observable object is no longer one
prompt and one response. The model makes a sequence of decisions under a
controller, invokes tools, observes results, and changes an external state.
This creates richer behavioral evidence, but it also introduces harness and
environment effects that can be mistaken for model identity.

Conversational-agent forensics infer base models and system prompts from
multi-turn text~\cite{white2026blackbox}. Closer to coding agents, ProcGrep
represents trajectories as programs and shows that procedural patterns carry
attributable signal; it also reports substantial same-model divergence across
harnesses~\cite{oderinwale2026procgrep}. AgentBench evaluates reasoning and
decision making in interactive environments~\cite{liu2024agentbench}, while
AgentDojo evaluates utility and prompt-injection robustness for tool-using
agents~\cite{debenedetti2024agentdojo}. These systems study forensics,
capability, safety, or procedural characterization rather than verification of
a remote model claim. \system{} instead elicits identity evidence with paired
coding tasks and records locally verifiable tool use, recovery behavior, and
repository consequences. It treats the harness as an explicit execution
context, maps native events from three CLI harnesses into the same behavioral
semantics, and excludes raw harness labels from the model fingerprint.

\section{Overview}
\label{sec:overview}

Figure~\ref{fig:architecture} maps five operational steps to the three design
components developed in Section~\ref{sec:design}. The verifier takes a claimed
identity, a coding-agent CLI, and a set of registered model identities. The
same behavioral semantics are instantiated on each supported CLI harness,
while enrollment and scoring remain conditioned on the harness that produced
the execution records.

\begin{enumerate}[leftmargin=*]
  \item \textbf{Probe execution and local records (Steps 1--2).} The
  verifier runs the A/B variants of the \textsc{Verify}, \textsc{Recover}, and
  \textsc{Resolve} probe pairs. Their six native trajectories form one fingerprint
  sample. The verifier records native events, repository outcomes, and
  controlled local checks, which are then used to construct behavioral
  features.

  \item \textbf{Feature extraction (Step 3).} A common representation maps
  harness-native records to shared action and outcome semantics.
  Instance-level features retain the decision structure of each probe
  variant, while distribution-level features summarize behavior across the
  complete sample.

  \item \textbf{Enrollment and model identification (Steps 4--5).} Repeated
  clean samples establish frozen feature processing and a regularized Gaussian
  mixture reference for each registered model under each harness. A later query is
  processed without updating those enrollment-fitted quantities and scored
  against every registered reference. The highest-scoring identity provides
  the candidate assignment; an open-set detector and calibrated threshold
  determine whether it is accepted or reported as unknown. Accepted identities
  are compared with the service's claim as specified in Section~\ref{sec:problem}.
\end{enumerate}

The small red callout and the green boundary in Figure~\ref{fig:architecture}
instantiate the threat model from Section~\ref{sec:problem}.
Provider-side routing, instructions, or output wrappers can affect a later
query fingerprint, whereas the CLI records local execution events and outcomes
on the user's host.
Sections~\ref{sec:probe-construction},
\ref{sec:feature-extraction}, and \ref{sec:model-identification} explain the
three design components; Sections~\ref{sec:setup} and \ref{sec:evaluation}
describe their instantiation and evaluation.

\section{Design}
\label{sec:design}

Figure~\ref{fig:architecture} outlines \system{}'s workflow, which comprises
three components: probe design, behavioral feature extraction, and model
identification. Three coding probe pairs elicit decisions about verification,
failure recovery, and conflicting requirements. The resulting trajectories
are encoded into instance-level features that preserve variant-specific
behavior and distribution-level features that summarize behavior across
the probe suite. A probabilistic identifier then compares each query
fingerprint with enrolled model references to infer model identity.

\subsubsection*{Design goals}
We organize \system{} around three goals that guide probe construction,
feature extraction, and model identification.

\textbf{G1: Expose coding-agent-specific behavior.}
The probes should elicit decisions about checking changes, recovering from
failures, and resolving conflicting feedback. Controlled task variants
should reveal how an agent adjusts its actions when these conditions change,
providing identity evidence beyond the final answer or patch.

\textbf{G2: Represent agent behavior consistently.}
The representation should preserve both decisions within individual probe
variants and behavioral patterns across the probe suite. Features should
have explicit behavioral meanings and retain those meanings across
supported harnesses, despite differences in tool names and event formats.

\textbf{G3: Identify models with limited enrollment data.}
The identifier should capture model-specific behavioral variation from a
small reference set while accounting for correlations among features.
Its scores should also support inspection of the behavioral evidence
behind an identity decision.

\subsection{Probe Design}
\label{sec:probe-construction}

\subsubsection*{From text outputs to agent behavior}
Conventional LLM fingerprints distinguish models using carefully chosen
prompts, response patterns, token probabilities, or output distributions.
These signals characterize model behavior under a direct prompt--response
interface. In a coding agent, however, the observed response also depends
on system instructions, controller logic, tool interfaces, and feedback
accumulated during execution. These factors can alter how model-specific
patterns appear in the final output. At the same time, agent execution
exposes additional behavioral choices: two agents may produce the same
correct patch while following different inspection, editing, testing,
and recovery procedures. The execution process therefore provides
identity evidence that the final answer or patch alone does not capture.

\system{} records action selection and ordering, responses to tool
feedback, and the final repository state. These observations describe
the execution process associated with each task outcome and form
the basis of the behavioral fingerprint.

\subsubsection*{Coding tasks as behavioral probes}
Code-modification tasks expose a sequence of decisions that are central to
coding-agent operation: inspecting the repository, locating relevant code,
editing files, choosing checks, interpreting execution feedback, and deciding
whether further changes are needed. Agents can reach the same task outcome
through different sequences of these decisions. Native tool events, test
results, and repository diffs make these differences observable, allowing
the verifier to examine both the execution process and the resulting state
without relying solely on the agent's account of its actions.

We construct each probe pair around a small repository task whose
instructions, initial files, tests, and expected outcome are controlled by
the verifier. Keeping the task compact limits unrelated implementation
choices, while controlled variants expose responses to a specific change
in the task conditions. This provides a common basis for comparing models
and repeating measurements under different harnesses. Prior work identifies
model-specific signals in agent procedures~\cite{oderinwale2026procgrep}.
Our probes examine verification, recovery, and conflict resolution through
controlled repository tasks, with observable action sequences and locally
checkable outcomes.

\subsubsection*{Three probe pairs}
We design three probe pairs around complementary decisions in coding-agent
execution: whether to verify a change, how to respond to an execution
failure, and which requirement to follow when instructions and tests
conflict. Each pair represents a common coding situation, permits multiple
action sequences, and produces locally checkable outcomes. The recorded
sequences distinguish how agents approach the task, including executions
that reach the same final outcome.

\begin{itemize}[leftmargin=*]
  \item \textsc{Verify} examines how the agent establishes that its change
  is correct. We record whether it performs a relevant check after the
  final code edit, which checks it chooses, and whether their results
  support its completion claim.

  \item \textsc{Recover} examines how the agent responds to a transient
  test failure. We record whether it retries immediately, inspects the
  cause, modifies code or execution commands, and obtains a successful
  result. The sequence of these actions distinguishes recovery strategies
  even when the final outcomes agree.

  \item \textsc{Resolve} examines how the agent handles a conflict between
  the user specification and a stale shipped test. We record whether its
  implementation follows the specification, whether it modifies the
  tests, and whether the final code passes a verifier-controlled check.
  These observations distinguish adherence to the requested behavior
  from attempts to satisfy conflicting test feedback.
\end{itemize}

The three pairs probe verification of the agent's own work, adaptation to
execution feedback, and resolution of conflicting requirements. They
sample distinct behavioral decisions without attempting to cover every
aspect of coding-agent operation. Their individual and combined
contributions are evaluated through the probe-pair ablations in
Section~\ref{sec:evaluation}.

\subsubsection*{Two variants reveal workflow changes}
Each probe pair contains two versions of the same task. Variant A provides
the baseline condition, while variant B changes a targeted instruction,
fixture, or execution response. Both variants retain the same requested
functionality and are executed in separate, fresh sessions. For example,
\textsc{Recover}-A allows the first test invocation to proceed normally,
whereas \textsc{Recover}-B introduces a controlled transient failure at
that invocation.

The paired design captures both the agent's baseline behavior and its
response to the changed condition. Observable differences include files
inspected, checks selected, retries, strategy changes, edit order, and the
final repository state. Holding the underlying task and requested outcome
constant reduces variation due to unrelated task content and makes these
differences easier to interpret. We also distinguish the condition
configured by the verifier from the feedback actually encountered by the
agent. In \textsc{Recover}, for example, an execution that never invokes
the affected test does not provide evidence of how the agent responds
to that failure.

We retain the ordered pair $(A,B)$ rather than only the difference $B-A$.
For a binary behavior, $(0,0)$ and $(1,1)$ have the same difference, although
the former indicates that the behavior occurs in neither variant and the
latter that it occurs in both. Keeping both observations distinguishes
whether a behavior occurs in neither variant, one variant, or both.

\subsubsection*{From trajectories to one fingerprint sample}
One execution of a probe variant produces a \emph{trajectory}.
Executing both variants of the three probe pairs yields six trajectories,
which together form one \emph{fingerprint sample}. This sample is the
unit used for feature extraction and model identification. Each
repetition executes the full probe suite in fresh sessions to produce
another fingerprint sample. Individual tool calls and constituent
trajectories are not counted as separate fingerprint samples.

\LidarTableFeatures

\subsection{Feature Extraction}
\label{sec:feature-extraction}

\subsubsection*{Feature space}
An execution trajectory records tool invocations, action order,
execution feedback, agent utterances, and the resulting repository state.
We encode these observations through behavioral measurements of
verification, recovery, task decisions, command use, and workflow
structure. The representation preserves both variant-specific
observations and statistical summaries across the probe suite.

The feature space has two complementary parts. Instance-level behavior
comprises 144 fields that preserve variant-specific task decisions,
verification actions, behavioral expression, and action-sequence structure.
Distribution-level behavior comprises 128 fields that summarize trajectory
descriptors across the six probe executions, including 68 mean statistics
and 60 standard-deviation statistics. Together, the two levels form a
272-dimensional representation that captures both local decisions and
behavioral patterns across the probe suite. The same feature definitions
are used across harnesses.

Model and provider identifiers, credentials, latency, and raw file paths
are excluded from the fingerprint. Commands and agent utterances are
processed into behavioral measurements rather than retained as raw
strings. The encoding distinguishes stated plans and reported outcomes
from actions and results recorded during execution, retaining their
association within each probe variant.

\subsubsection*{Harness event normalization}
OpenCode, OpenHands, and mini-swe-agent expose execution records through
different event formats and tool interfaces. We map native records to
shared action roles, including inspection, file modification, successful
and failed execution, and agent communication. The mapping considers
tool arguments and execution results as well as tool names: a shell
invocation, for example, may inspect a file, modify it, or run a test.
The resulting action roles have common behavioral meanings across
the supported harness interfaces.

The normalized records retain action order, execution outcomes, and the
association between agent statements and observable actions. Feature
extraction then applies common rules to these records, producing the same
feature definitions and dimensionality across harnesses. Model references
and identifiers are fitted separately for each harness. Numerical
missing-value handling and scaling are determined from enrollment data,
as described in Section~\ref{sec:scoring}.

\subsubsection*{Instance-level behavior}
The 144 instance-level fields remain associated with specific probe
variants. They comprise 23 task-state fields, 47 verification-detail
fields, 58 behavioral-expression fields, and 16 action-structure fields.
Task-state fields describe decisions such as whether an edit remains
unverified, whether a transient failure was encountered, and whether the
agent inspected its cause or retried without inspection. Verification
details record tool use, check invocations, execution outcomes, and the
ordering of checks relative to edits.

Behavioral-expression fields distinguish plans, reports, and mentions
of verification, recovery, and conflict handling in intermediate and
final agent utterances. Where applicable, these statements are related
to observed check outcomes, distinguishing, for example, a successful-test
report accompanied by a successful check from one without an observed
check. Action-structure fields describe transitions, repeated workflow
stages, and patterns such as execution after modification.

The association with probe variants preserves distinctions that aggregate
counts do not identify. Equal test counts, for example, can correspond
to different verification behavior depending on whether testing occurs
before or after the final edit.

\subsubsection*{Distribution-level behavior}
The 128 distribution-level fields summarize trajectory descriptors across
the six executions. These descriptors cover action frequency, command
use, reading--editing--testing order, repetition, recovery, task outcomes,
and response form. The representation retains 68 mean statistics and
60 sample-standard-deviation statistics. Let $X_j$ contain the trajectories
in fingerprint sample $X$ for which descriptor $g_j$ is defined. With
$\mathcal{M}$ and $\mathcal{S}$ denoting the retained mean and
standard-deviation descriptor sets, respectively,
\[
\begin{aligned}
\phi_{\mathrm{dist}}(X)
 &= \bigl[\mu_{T\in X_j}g_j(T)\bigr]_{j\in\mathcal{M}}
 \oplus \bigl[s_{T\in X_j}g_j(T)\bigr]_{j\in\mathcal{S}},\\
|\mathcal{M}|&=68,\qquad |\mathcal{S}|=60,
\end{aligned}
\]
where $\oplus$ denotes concatenation. The mean describes the average
level of a trajectory descriptor, while the sample standard deviation
describes its variation across probe conditions within the same
fingerprint sample. An undefined descriptor is excluded from the corresponding
summary; a mean requires at least one valid observation, and a sample
standard deviation requires at least two. Statistics that cannot be
computed remain missing until enrollment-based preprocessing.

The combined representation retains variant-specific measurements
alongside their statistical summaries across the probe suite:
\[
\phi(X)=\phi_{\mathrm{inst}}(X)\oplus\phi_{\mathrm{dist}}(X)
\in\mathbb{R}^{144+128}=\mathbb{R}^{272}.
\]
Dependencies among these measurements are incorporated into the
reference distribution used for identification in Section~\ref{sec:scoring}.

\LidarTableProbes

\subsection{Model Identification}
\label{sec:model-identification}
\label{sec:scoring}

Given a query fingerprint, \system{} identifies the registered model
whose reference distribution best matches the observed behavior.
For model $c$ under harness $H$, the enrollment set
$\mathcal{E}_{c,H}$ contains $n_c$ fingerprint samples, each represented
by the 272 features defined in Section~\ref{sec:feature-extraction}.
A sample consists of six probe trajectories; its constituent tool
calls and trajectories are not treated as separate enrollment samples.
Identifiers are fitted separately for each harness. We omit $H$
from the following notation.

The feature representation contains correlated measurements of agent
execution, including action frequencies, verification outcomes, and
workflow transitions. Repeated executions also exhibit within-model
variation, as the same model may follow different action sequences
under the same task conditions. We model these observations with
reference-centered Gaussian components sharing a regularized
covariance matrix. This formulation represents both dependencies
among features and variation among reference executions.

\subsubsection*{Feature preprocessing}
We replace non-finite feature values with their enrollment medians,
using zero for fields without finite enrollment observations.
Each field is then centered by its enrollment mean and scaled by
its enrollment standard deviation. A scale of one is used for
fields whose standard deviation is at most $10^{-12}$.
The resulting transformation retains all 272 coordinates, including
fields that are constant in the enrollment set.

Preprocessing parameters are estimated from enrollment data and
held fixed during query evaluation. Indicators, counts, proportions,
and continuous summaries are represented numerically without
discretization, preserving their observed variation.

\subsubsection*{Reference distribution estimation}
Let $\mathbf{x}_{ci}$ denote the standardized vector of reference
sample $i$ for model $c$, and let $\boldsymbol{\mu}_c$ denote the
mean of its reference vectors. We estimate a shared within-model
covariance matrix $\widehat{\Sigma}$ by pooling class covariance
estimates with automatic shrinkage. A small diagonal regularizer
is added for numerical stability. The shared covariance captures
feature dependencies using observations from all registered
models, while avoiding a separate unrestricted covariance estimate
from each model's limited reference set.

Each reference vector defines a Gaussian component centered at
\begin{equation}
 \mathbf{m}_{ci}
 = \boldsymbol{\mu}_c
 + \lambda(\mathbf{x}_{ci}-\boldsymbol{\mu}_c),
 \qquad 0\leq\lambda\leq1.
 \label{eq:reference-centers}
\end{equation}
Here, $\lambda$ controls the dispersion of component centers around
the model mean. Setting $\lambda=0$ yields a single center at
$\boldsymbol{\mu}_c$, whereas $\lambda=1$ retains the reference
observations as component centers. Intermediate values shrink
sample-specific deviations while preserving variation among
reference executions.

The reference distribution is an equally weighted mixture of
these components, each with covariance $t\widehat{\Sigma}$,
where $t>0$ determines their spread. The center-shrinkage parameter
$\lambda$ and covariance scale $t$ control the flexibility of
the distribution within a common model family.

\LidarTableModels

\subsubsection*{Query scoring and identification}
For a standardized query vector $\mathbf{x}$, its squared
Mahalanobis distance to reference component $i$ of model $c$ is
\[
 d_{ci}^{2}(\mathbf{x})
 =(\mathbf{x}-\mathbf{m}_{ci})^\top
   \widehat{\Sigma}^{-1}
   (\mathbf{x}-\mathbf{m}_{ci}).
\]
The covariance-weighted distance measures deviations relative
to the within-model variability of the features, accounting
for correlations between their coordinates.

The query score is the log mixture likelihood with the common
Gaussian normalization term omitted:
\begin{equation}
 s_c(\mathbf{x})
 = \log\!\left[
 \frac{1}{n_c}\sum_{i=1}^{n_c}
 \exp\!\left(-\frac{d_{ci}^{2}(\mathbf{x})}{2t}\right)
 \right].
 \label{eq:model-score}
\end{equation}
The omitted term is identical across registered models within
a harness and therefore does not affect their ranking.
Each reference component contributes according to its proximity
to the query, and the factor $1/n_c$ normalizes for the number
of components.

Under equal model priors, the closed-set decision is
\begin{equation}
 c^*(\mathbf{x})=\arg\max_c s_c(\mathbf{x}).
 \label{eq:closed-set-decision}
\end{equation}
At $\lambda=0$, this rule reduces to Gaussian discriminant
classification with shared covariance and equal priors.
For $\lambda>0$, the score incorporates the distribution of
reference-centered components rather than only the model mean.

We select $\lambda$ and $t$ through validation on enrollment
rounds. The OpenCode and OpenHands experiments use the same
selected configuration, $\lambda=0.25$ and $t=4$, with reference
means, component centers, and covariance estimates fitted
separately for each harness. Each query fingerprint is scored
individually using the fitted identifier.

The scores express relative likelihoods under the reference
distributions, rather than calibrated identity probabilities.
Their quadratic terms and mixture weights provide the basis
for the score-margin decomposition used in the behavioral
evidence analysis in Section~\ref{sec:evaluation}.

\LidarTableSampling

\subsubsection*{Open-set identification}
A closed-set identifier assigns every query to a registered model,
including queries generated by models absent from the reference library.
Open-set identification additionally determines whether the query is
consistent with the registered set. We separate these decisions:
the identity classifier selects the highest-scoring registered model,
and an unknown detector assesses whether to accept that assignment.

For query $\mathbf{x}$, let $c^*(\mathbf{x})$ denote the identity selected
by Equation~\ref{eq:closed-set-decision}. We construct a detector input
$\boldsymbol{\psi}(\mathbf{x})$ from summaries of the reference-matching
scores and distances. These summaries describe the strength of the
leading match, its separation from competing identities, and the query's
proximity to the reference distributions. The detector therefore
considers both absolute compatibility with the reference library and
relative separation among candidate identities.

A logistic detector maps these summaries to a membership score:
\begin{equation}
 q(\mathbf{x})
 = \sigma\!\left(
 \mathbf{w}^{\top}\boldsymbol{\psi}(\mathbf{x})+b
 \right),
 \label{eq:membership-score}
\end{equation}
where $\sigma$ is the logistic function. The detector is fitted using
registered-model samples as positive examples and development-unknown
samples as negative examples. Development unknowns represent identities
outside the reference library and are distinct from the unknown
identities reserved for final evaluation. The identity classifier
remains fixed during detector fitting.

A separate calibration subset determines the acceptance threshold
$\theta$. Calibration evaluates incorrect acceptance of unknown queries
and incorrect identity assignments to registered queries. The resulting
open-set decision is
\begin{equation}
 \widehat c(\mathbf{x})=
 \begin{cases}
 c^*(\mathbf{x}), & q(\mathbf{x})\geq\theta,\\
 \textsc{Unknown}, & q(\mathbf{x})<\theta.
 \end{cases}
 \label{eq:open-set-decision}
\end{equation}
The detector determines whether the leading identity is accepted;
it does not change the ranking of registered candidates. A registered
query is correctly identified only when its identity assignment is
correct and its membership score meets the threshold.

Reference fitting, detector development, and threshold calibration
use separate data partitions. Final unknown identities participate
in none of these stages. After calibration, the identity classifier,
detector, and threshold remain fixed during query evaluation.
Section~\ref{sec:openset-current} specifies the identity partition
and calibration procedure used in the experiments.

\section{Evaluation}
\label{sec:evaluation}

We evaluate \system{} on 24 model identities from eight families under three
coding-agent harnesses. We first describe the experimental setup and compare
closed-set identification performance against three baselines. We then
evaluate \system{} in an open-set setting that includes unregistered models.
Finally, we conduct ablation studies and a supplementary evaluation of
robustness to provider-side manipulation.

\subsection{Experimental Setup}
\label{sec:setup}

\subsubsection*{Probe suite and sampling unit}
Table~\ref{tab:probes} specifies the three probe pairs. A fingerprint consists
of one execution of each of the six variants. Every variant starts a fresh
session and an isolated fixture copy, without shared conversation history or
persistent memory. The six trajectories form one statistical sample, they are not
six independent identity decisions.

\LidarTableClosedSet

\subsubsection*{Models and Harnesses}
Table~\ref{tab:current-model-panel} lists the evaluated models.

We use three established open-source coding-agent harnesses:
OpenCode~1.19.4~\cite{opencode_current},
OpenHands SDK~1.58.1~\cite{openhands_sdk}, and
mini-swe-agent~2.4~\cite{minisweagent}.
They support multiple model providers and offer different tool interfaces
and execution workflows.
Together, they represent three common harness designs: an interactive
coding CLI (OpenCode), a programmable agent SDK (OpenHands), and a
lightweight scaffold for software-engineering benchmarks (mini-swe-agent).
We choose mini-swe-agent over SWE-agent following the maintainers'
recommendation, which cites its simpler design and comparable
performance~\cite{sweagent_repository}.

We execute the same probe suite through each harness's native tools and
construct references and identifiers separately for each harness.
Model identifiers are hidden from the agent, and system-prompt selection
is kept model-independent within each harness.

We use OpenCode and OpenHands for method development, selecting a common
feature representation and classifier configuration for both harnesses.
We then evaluate mini-swe-agent with the same feature definitions and
hyperparameters, without further configuration tuning. Reference
distributions are fitted separately using each harness's enrollment data.
The mini-swe-agent evaluation therefore examines the applicability of
the selected method to an additional agent implementation beyond those
used for configuration selection.

For each model--harness pair, we repeat the probe suite across multiple
collection rounds, using a fresh session for each probe variant. Reference,
calibration, and test fingerprints are collected in separate executions, and
evaluation metrics aggregate results across repeated test rounds.
Reference data are used to fit the identifier.
Calibration data are split between developing and fitting the unknown-model
detector and setting its decision threshold. Test data are used for evaluation. Generation settings are fixed for each model according to its supported
controls. Reference requests omit seeds; where supported, calibration
and test requests use disjoint seed ranges.

\subsubsection*{Baselines}
We compare \system{} with LLMmap~\cite{pasquini2025llmmap}, Model Equality
Testing (MET)~\cite{gao2025modelequality}, and
ZeroPrint~\cite{shao2026zeroprint} under the same harness versions.
We keep the three baselines' sampling configurations within a comparable order of
magnitude while retaining their core fingerprinting procedures.
Table~\ref{tab:sampling-budget} summarizes the sampling configurations.

LLMmap identifies models from their responses to a fixed set of probes using
a learned encoder and classifier. MET compares response distributions using
Hamming--MMD; we adapt it to identification by matching each query to the
closest registered reference. ZeroPrint characterizes responses to perturbed
queries using ridge regression and matches the resulting fingerprints by
Pearson correlation. Protocol adaptations and collection exceptions are
recorded in the artifact.

\subsubsection*{Evaluation Metrics}
We evaluate closed-set identification using Top-$k$ accuracy and mean
reciprocal rank (MRR). Let $r_i$ denote the rank of the true identity for
query $i$. For $k\in\{1,3\}$,
\begin{equation}
 \mathrm{Top}\text{-}k=\frac{\sum_{i=1}^{N}\mathbf{1}[r_i\leq k]}{N},
 \qquad
 \mathrm{MRR}=\frac{\sum_{i=1}^{N}1/r_i}{N}.
\label{eq:topk}
\end{equation}
For open-set evaluation, let $\mathcal K$ and $\mathcal U$ denote registered
and unregistered test samples. Let $\hat y_i$ be the predicted identity and
$a_i=a_i(\tau)$ indicate acceptance at threshold $\tau$. Known correct
acceptance (KCA) and unknown false acceptance (FAR) are
\begin{equation}
 \mathrm{KCA}=\frac{\sum_{i\in\mathcal K}\mathbf{1}[\hat y_i=y_i]\,a_i}{|\mathcal K|},
 \qquad
 \mathrm{FAR}=\frac{\sum_{i\in\mathcal U}a_i}{|\mathcal U|}.
\label{eq:kca-far}
\end{equation}
We summarize performance at the calibrated threshold using balanced
open-set accuracy, $\mathrm{BAcc}=(\mathrm{KCA}+1-\mathrm{FAR})/2$.
We additionally report OSCR to summarize the trade-off between KCA and FAR
across thresholds.

\newcommand{\evaltakeaway}[1]{%
  \begin{tcolorbox}[
    colback=black!3,
    colframe=black!45,
    boxrule=0.6pt,
    arc=1.5mm,
    outer arc=1.5mm,
    boxsep=0pt,
    left=5.5pt,
    right=5.5pt,
    top=5pt,
    bottom=5pt,
    before skip=0.7\baselineskip,
    after skip=0.7\baselineskip]
    \textbf{Takeaway.} #1
  \end{tcolorbox}%
}

\subsection{Closed-Set Identification}
\label{sec:closed-current}

We first evaluate whether \system{} can identify the underlying model from
24 registered identities. Table~\ref{tab:overall-identification} summarizes
the results under the three harnesses. \system{} achieves Top-1 accuracy
between 91.39\% and 98.89\%, consistently identifying the correct model in
each setting. Its high Top-3 accuracy and MRR further show that the true
identity remains near the top of the candidate ranking even when the
first-ranked prediction is incorrect.

\LidarTableOpenSet

\system{} outperforms all three baselines on Top-1, Top-3, and MRR under
each harness, with an average Top-1 improvement of over 20 percentage points
over the best baseline in each setting. The advantage also holds over LLMmap and ZeroPrint, whose relative
performance varies with the harness. Together, these comparisons support
the effectiveness of \system{} for identifying models through coding-agent
execution, with a consistent advantage over the evaluated fingerprinting
baselines.

\LidarTableAblation

These results demonstrate that \system{} is applicable to different types
of coding-agent harnesses, spanning an interactive coding CLI, a
programmable agent SDK, and a lightweight software-engineering scaffold.
Its consistently strong performance across these settings supports the
broader applicability of behavioral fingerprinting for model identification
in coding agents.

\evaltakeaway{Across three types of coding-agent harnesses, \system{} achieves
over 95\% average closed-set identification accuracy, improving on the
baselines by over 20 percentage points on average. These results support the
applicability of behavioral fingerprinting across different harness designs.}

\subsection{Open-Set Identification}
\label{sec:openset-current}

We evaluate whether \system{} can identify registered models while
distinguishing models absent from its reference library. We partition the
24 identities into 16 registered models, four development unknowns, and four
test unknowns, using the same partition for OpenCode and OpenHands.
The registered set contains two models from each of the eight families.
For each of the four families with additional identities---GPT, Claude,
Qwen, and GLM---we reserve one identity for development and another for
testing. This preserves family
diversity in the reference library while testing unfamiliar identities
from families already represented.

\system{} makes two decisions: which registered model best matches a query,
and whether the query belongs to the registered set. The identity classifier
is trained on reference data from registered models. To develop the unknown
detector, we use calibration data from both registered models and the four
development unknowns, providing examples of identities inside and outside
the reference library. Nine calibration groups per model are used to develop
and fit this detector; six separate groups determine its decision threshold.

The four test unknowns are reserved for final evaluation and participate in
neither training nor calibration. This tests the detector on unknown
identities different from those used during development. LLMmap uses the
same registered and test-unknown identities; its reproduced distance-based
detector is trained without the four development unknowns.

Table~\ref{tab:openset-identification} reports the comparison.
\system{} improves known correct acceptance and reduces unknown false
acceptance under harnesses relative to the current LLMmap reproduction.
Its balanced open-set accuracy is approximately 80\% in these settings,
combining correct identification of registered models with discrimination
of unregistered identities. The higher OSCR values also indicate a better
trade-off across thresholds.

Open-set decisions must balance identifying registered models against
avoiding incorrect assignments to unregistered models. Under the selected
thresholds, \system{} correctly identifies and accepts more than 70\% of
registered queries. The remaining registered queries are either rejected or accepted under an
incorrect identity.

\evaltakeaway{\system{} distinguishes unregistered identities from represented
model families while retaining useful identification of registered models,
achieving a stronger open-set trade-off than LLMmap under three harnesses.}

\subsection{Ablation Study}
\LidarFigClassifierAblation

We examine feature levels, probe design, reference size, calibration
allocation, and classifier choice. We fit models separately on OpenCode
and OpenHands and average their results. These two harnesses let us test
whether the benefits of the feature and probe design persist across agent
implementations. Within each experiment, we vary
one component, refit the model, and use the same test queries.\par
\textbf{Feature complementarity.}
Using both feature levels yields 95.14\% Top-1 accuracy
(Table~\ref{tab:ablation}). Accuracy drops to 93.47\% with only
distribution-level features and to 87.08\% with only instance-level
features. With instance-level features fixed, using only means or only standard
deviations lowers accuracy: both average behavior and variation matter.\par
\textbf{Probe contributions.}
Each probe pair contributes to identification. Removing \textsc{Resolve}
produces the largest reduction, at 6.94 percentage points, while removing
\textsc{Verify} or \textsc{Recover} also lowers accuracy. Neither variant
alone matches the complete suite: A-only and B-only configurations achieve
87.92\% and 90.56\%, respectively. These results support retaining the three
behavioral mechanisms and both variants. Because variant removal also reduces
the number of observed executions, this comparison does not isolate the
benefit of pairing at a fixed sampling count.\par
\textbf{Reference sample size.}
Most of the improvement occurs within the first nine reference groups
(Figure~\ref{fig:classifier-ablation}(a)). Nine groups achieve 93.99\%
mean Top-1, compared with 95.14\% using all fifteen groups; increasing to
twelve groups narrows the difference to less than half a percentage point.
The feature set and classifier configuration were selected with fifteen
reference groups and remain fixed, while reference-dependent parameters are
refitted on each subset. The error bands summarize variation across ten
shared reference subsets.\par
\textbf{Calibration allocation.}
Changing the split between gate fitting and threshold setting has little
effect on balanced accuracy but changes which queries are accepted
(Figure~\ref{fig:classifier-ablation}(b)). The 12/3 split lowers FAR to
2.50\% while reducing KCA to 58.75\%, illustrating the trade-off between rejecting unknown queries
and accepting correctly identified registered models.\par
\textbf{Classifier choice.}
We compare eight classifiers using the same behavioral features and
reference/query split (Figure~\ref{fig:classifier-ablation}(c)). The multimode Gaussian
classifier achieves the highest mean Top-1 accuracy among the evaluated
configurations, at 95.14\%, followed closely by shrinkage LDA at 94.58\%.
Logistic regression reaches 92.50\%, whereas simpler prototype and
independent-feature classifiers perform less well. The modest gain over shrinkage LDA suggests a benefit from modeling
multiple reference patterns.

\evaltakeaway{With both feature levels and all probes, the multimode Gaussian
classifier reaches 93.99\% mean Top-1 accuracy with nine reference groups
and 95.14\% with fifteen.}

\subsection{Robustness to Provider-Side Manipulation}
\label{sec:robustness}

We evaluate whether \system{} remains effective when explicit identity
cues or response presentation are altered. We test identity obfuscation
through system instructions (A1) and constrained transformations
of final responses (A2).\par
\textbf{Setup.}
We conduct a supplementary evaluation on six models:
GLM-5.1, GLM-5.2, GLM-5.3, Qwen3.6-Plus, Qwen3.7-Plus, and Qwen3.8-27B.
LLMmap, MET, and
ZeroPrint retain their native representations and scoring procedures,
with clean data and separately collected clean reference
responses.\par
\textbf{Manipulations.}
For \emph{A1 (identity obfuscation)}, we use a
provider-side system instruction to suppress or alter explicit model
identity claims. For \emph{A2 (constrained response transformation)}, we apply
a fixed transformation to the final natural-language response.
A2 representations are recomputed from the transformed
responses.\par
\LidarFigBehaviorCase
\LidarFigFeatureImportance
\textbf{Evaluation.}
We compare each method with its own clean-query result using Top-1 accuracy
and MRR. These manipulations test identity obfuscation and response
formatting; they do not establish resistance to targeted behavioral
imitation or selective routing of audit tasks.\par
\newcommand{\LidarRobustnessTakeaway}{%
  \evaltakeaway{On the six-model panel, \system{} retains 100\% Top-1 accuracy
  on OpenCode under identity obfuscation and drops from 97.22\% to 91.67\% on
  mini-swe-agent. Under response transformation, it retains its clean accuracy
  and MRR on both harnesses.}}
\textbf{Results.}
Table~\ref{tab:robustness} in Appendix~\ref{app:robustness} reports the results.
Under A1, \system{} retains
the highest Top-1 accuracy on OpenCode; under A2, its
Top-1 accuracy and MRR remain unchanged on both harnesses.
Under A1, Top-1 accuracy on mini-swe-agent drops by 6.67 percentage points
for LLMmap and 33.33 points for both MET and ZeroPrint. Under A2, MET drops
by 66.67 and 33.33 points on OpenCode and mini-swe-agent, respectively.

\LidarRobustnessTakeaway

\section{Discussion}
\label{sec:discussion}

\begingroup
\makeatletter
\def\@IEEEsectpunct{\ }
\renewcommand{\paragraph}{%
  \@startsection{paragraph}{4}{0pt}%
    {1.5ex plus 0.2ex minus 0.2ex}{-1em}%
    {\normalfont\normalsize\bfseries}}
\makeatother

\begingroup
\makeatletter
\def\@IEEEsectpunct{\ }
\renewcommand{\paragraph}{%
  \@startsection{paragraph}{4}{0pt}%
    {1.5ex plus 0.2ex minus 0.2ex}{-0.8em}%
    {\normalfont\normalsize\bfseries}}
\makeatother
\subsection{What Makes an Agentic Fingerprint?}
\label{sec:feature-importance}

We use the closed-set data and fitted identifiers to examine how identity
evidence is distributed across behaviors. For each query, we decompose the
score difference between the true identity and its highest-scoring rival
into signed feature contributions and a separate mixture-weighting term.

We group features measuring the same behavior across probe variants or
statistical summaries and take the absolute value of each group's summed
contributions. To assess consistency, we compare three consecutive blocks
of five test rounds (B1--B3), each covering all evaluated models under both
harnesses.

\subsubsection*{Illustrative trajectories.}
Figure~\ref{fig:behavior-case-study} contrasts three illustrative executions
of the same whitespace-handling task under OpenCode. DeepSeek V4-Pro inspects
and edits the file before replying, whereas Kimi K2.7 re-reads the modified
file before concluding. MiniMax M3 interleaves explanations with its actions,
inspects related usages and tests, and successfully runs the unit tests.
These traces differ in action ordering, post-edit verification, and the
placement of visible explanations, exposing behavioral differences beyond
the final code change. The shared task context makes these differences
directly comparable across models.

\textbf{Evidence composition.} Trajectory organization accounts for about
28\% of total absolute evidence under both harnesses
(Figure~\ref{fig:feature-importance-story}(a)). Command use, task decisions,
and visible expression also contribute. Their relative shares remain similar
across the three test subsets within each harness, although response and
expression account for a larger share under OpenCode than under OpenHands.\par
\textbf{Mechanism consistency.} Ten feature groups rank highly under both
harnesses, including verification plans and mentions, command choice,
file inspection, edit-to-test transitions, and response characteristics
(Figure~\ref{fig:feature-importance-story}(b)). Several recur in all three
subsets. The heatmap gives each group's top-20\%
frequency among queries in each subset. Model coverage is the smaller of the
two harness-specific fractions of identities with at least one such occurrence.\par
\textbf{Evidence concentration.} For each correctly identified query, we
rank feature groups by absolute contribution, excluding the mixture-weighting
term (Figure~\ref{fig:feature-importance-story}(c)). The largest group
accounts for a median of 12.2\% of feature evidence under OpenCode and
10.0\% under OpenHands. Reaching 80\% of the evidence requires a median of
25 groups under OpenCode and 31 under OpenHands.

\endgroup 
\endgroup

\subsection{Limitations and Future Work}
\label{sec:limitations}

\textbf{Generalization across agent settings.}
Our evaluation focuses on coding agents with known, locally controlled
harnesses, using separate behavioral references for each harness.
Identification through unknown harnesses remains untested.
Future work should explore behavioral signals that generalize across
tools and controllers, extending \system{} to broader agent settings
and encapsulated services whose internal harness is unknown.\par
\textbf{Model identification in multi-model agents.}
Our evaluation considers a single model identity for each fingerprint.
Future work will explore agents that route different execution steps
to different models, investigating whether behavioral traces can
reveal model switches and attribute individual stages to the
contributing models.\par
\textbf{Robustness to behavioral imitation.}
Beyond the identity obfuscation and response transformation evaluated in
Section~\ref{sec:robustness}, future work should examine targeted imitation
of a registered model through fine-tuning or provider-side instructions.
Such studies should measure false acceptance of the target identity
alongside task performance and investigate whether more diverse probes
improve robustness.

\section{Conclusion}

\system{} checks model-identity claims through the coding-agent interface
by observing how agents verify changes, recover from failures, and resolve
specification--test conflicts. Across three harnesses and 24 models from
eight families, \system{} achieves over 95\% average closed-set identification
accuracy and approximately 80\% balanced \mbox{open-set accuracy}.

\bibliographystyle{IEEEtran}
\bibliography{references}

@inproceedings{pasquini2025llmmap,
  author = {Dario Pasquini and Evgenios M. Kornaropoulos and Giuseppe Ateniese},
  title = {{LLMmap}: Fingerprinting for Large Language Models},
  booktitle = {34th USENIX Security Symposium (USENIX Security 25)},
  year = {2025},
  pages = {299--318},
  publisher = {USENIX Association},
  note = {\url{https://www.usenix.org/conference/usenixsecurity25/presentation/pasquini}}
}

@inproceedings{gao2025modelequality,
  author = {Irena Gao and Percy Liang and Carlos Guestrin},
  title = {Model Equality Testing: Which Model Is This {API} Serving?},
  booktitle = {The Thirteenth International Conference on Learning
    Representations (ICLR)},
  year = {2025},
  note = {\url{https://openreview.net/forum?id=QCDdI7X3f9}}
}

@inproceedings{chen2022modelchange,
  author = {Lingjiao Chen and Matei Zaharia and James Y. Zou},
  title = {How Did the Model Change? {Efficiently} Assessing Machine Learning
    {API} Shifts},
  booktitle = {The Tenth International Conference on Learning
    Representations (ICLR)},
  year = {2022},
  note = {\url{https://people.eecs.berkeley.edu/~matei/papers/2022/iclr_masa.pdf}}
}

@inproceedings{shao2026zeroprint,
  author = {Shuo Shao and Yiming Li and Hongwei Yao and Yifei Chen and Yuchen Yang and Zhan Qin},
  title = {Reading Between the Lines: Towards Reliable Black-box {LLM} Fingerprinting via Zeroth-order Gradient Estimation},
  booktitle = {Proceedings of the ACM Web Conference 2026},
  year = {2026},
  pages = {2637--2648},
  doi = {10.1145/3774904.3792196},
  note = {\url{https://doi.org/10.1145/3774904.3792196}}
}

@inproceedings{gubri2024trap,
  author = {Martin Gubri and Dennis Ulmer and Hwaran Lee and Sangdoo Yun and
    Seong Joon Oh},
  title = {{TRAP}: Targeted Random Adversarial Prompt Honeypot for Black-Box
    Identification},
  booktitle = {Findings of the Association for Computational Linguistics:
    ACL 2024},
  year = {2024},
  pages = {11496--11517},
  publisher = {Association for Computational Linguistics},
  doi = {10.18653/v1/2024.findings-acl.683},
  note = {\url{https://aclanthology.org/2024.findings-acl.683/}}
}

@misc{ren2025cotsrf,
  author = {Zhenzhen Ren and GuoBiao Li and Sheng Li and Zhenxing Qian and
    Xinpeng Zhang},
  title = {{CoTSRF}: Utilize Chain of Thought as Stealthy and Robust
    Fingerprint of Large Language Models},
  year = {2025},
  howpublished = {arXiv preprint arXiv:2505.16785},
  eprint = {2505.16785},
  archivePrefix = {arXiv},
  primaryClass = {cs.CR},
  note = {\url{https://arxiv.org/abs/2505.16785}}
}

@misc{bruckner2026onetoken,
  author = {Tom{\'a}{\v{s}} Bruckner},
  title = {One Token Is Enough: Fingerprinting and Verifying Large Language Models from Single-Token Output Distributions},
  year = {2026},
  howpublished = {arXiv preprint arXiv:2607.10252},
  eprint = {2607.10252},
  archivePrefix = {arXiv},
  primaryClass = {cs.CR},
  note = {\url{https://arxiv.org/abs/2607.10252}}
}

@misc{zhang2026ghostprint,
  author = {Jiahao Zhang and Xiuyu Li and Suhang Wang},
  title = {Your `{Pro}' {LLM} Subscription May Actually Be `{Free}': Exposing Fingerprint Spoofing Risks in {LLM} Inference Services},
  year = {2026},
  howpublished = {arXiv preprint arXiv:2606.16100},
  eprint = {2606.16100},
  archivePrefix = {arXiv},
  primaryClass = {cs.CR},
  note = {\url{https://arxiv.org/abs/2606.16100}}
}

@misc{white2026blackbox,
  author = {Isadora White and Yasaman Jafari and Taylor Berg-Kirkpatrick},
  title = {Black-Box Forensics for Conversational {LLM} Agents},
  year = {2026},
  howpublished = {arXiv preprint arXiv:2606.22698},
  eprint = {2606.22698},
  archivePrefix = {arXiv},
  primaryClass = {cs.CR},
  note = {\url{https://arxiv.org/abs/2606.22698}}
}

@inproceedings{hu2026llmprint,
  author = {Yuepeng Hu and Zhengyuan Jiang and Mengyuan Li and Osama Ahmed and
    Zhicong Huang and Cheng Hong and Neil Zhenqiang Gong},
  title = {Fingerprinting {LLM}s via Prompt Injection},
  booktitle = {Proceedings of the 64th Annual Meeting of the Association for
    Computational Linguistics (Volume 1: Long Papers)},
  year = {2026},
  pages = {11795--11810},
  publisher = {Association for Computational Linguistics},
  doi = {10.18653/v1/2026.acl-long.541},
  note = {\url{https://aclanthology.org/2026.acl-long.541/}}
}

@inproceedings{xu2024instructional,
  author = {Jiashu Xu and Fei Wang and Mingyu Ma and Pang Wei Koh and
    Chaowei Xiao and Muhao Chen},
  title = {Instructional Fingerprinting of Large Language Models},
  booktitle = {Proceedings of the 2024 Conference of the North American
    Chapter of the Association for Computational Linguistics: Human Language
    Technologies (Volume 1: Long Papers)},
  year = {2024},
  pages = {3277--3306},
  publisher = {Association for Computational Linguistics},
  doi = {10.18653/v1/2024.naacl-long.180},
  note = {\url{https://aclanthology.org/2024.naacl-long.180/}}
}

@misc{shao2025sok,
  author = {Shuo Shao and Yiming Li and Yu He and Hongwei Yao and Wenyuan Yang
    and Dacheng Tao and Zhan Qin},
  title = {{SoK}: Large Language Model Copyright Auditing via Fingerprinting},
  year = {2025},
  howpublished = {arXiv preprint arXiv:2508.19843},
  eprint = {2508.19843},
  archivePrefix = {arXiv},
  primaryClass = {cs.CR},
  note = {\url{https://arxiv.org/abs/2508.19843}}
}

@misc{zhu2025rut,
  author = {Xiaoyuan Zhu and Yaowen Ye and Tianyi Qiu and Hanlin Zhu and
    Sijun Tan and Ajraf Mannan and Jonathan Michala and Raluca Ada Popa and
    Willie Neiswanger},
  title = {Auditing Black-Box {LLM} {API}s with a Rank-Based Uniformity Test},
  year = {2025},
  howpublished = {arXiv preprint arXiv:2506.06975},
  eprint = {2506.06975},
  archivePrefix = {arXiv},
  primaryClass = {cs.CR},
  note = {\url{https://arxiv.org/abs/2506.06975}}
}

@misc{cai2025payfor,
  author = {Will Cai and Tianneng Shi and Xuandong Zhao and Dawn Song},
  title = {Are You Getting What You Pay For? {Auditing} Model Substitution in
    {LLM} {API}s},
  year = {2025},
  howpublished = {arXiv preprint arXiv:2504.04715},
  eprint = {2504.04715},
  archivePrefix = {arXiv},
  primaryClass = {cs.CL},
  note = {\url{https://arxiv.org/abs/2504.04715}}
}

@misc{oderinwale2026procgrep,
  author = {Hamidah Oderinwale},
  title = {Agent Trajectories as Programs: Fingerprinting and Programming
    Coding-Agent Behavior},
  year = {2026},
  howpublished = {arXiv preprint arXiv:2606.16988},
  eprint = {2606.16988},
  archivePrefix = {arXiv},
  primaryClass = {cs.SE},
  note = {\url{https://arxiv.org/abs/2606.16988}}
}

@misc{huang2025agentguide,
  author = {Kaibo Huang and Zipei Zhang and Zhongliang Yang and Linna Zhou},
  title = {{Agent Guide}: A Simple Agent Behavioral Watermarking Framework},
  year = {2025},
  howpublished = {arXiv preprint arXiv:2504.05871},
  eprint = {2504.05871},
  archivePrefix = {arXiv},
  primaryClass = {cs.AI},
  doi = {10.48550/arXiv.2504.05871},
  note = {\url{https://arxiv.org/abs/2504.05871}}
}

@inproceedings{huang2026agentmark,
  author = {Kaibo Huang and Jin Tan and Yukun Wei and Wanling Li and Zipei Zhang
    and Hui Tian and Zhongliang Yang and Linna Zhou},
  title = {{AgentMark}: Utility-Preserving Behavioral Watermarking for Agents},
  booktitle = {Proceedings of the 64th Annual Meeting of the Association for
    Computational Linguistics (Volume 1: Long Papers)},
  year = {2026},
  pages = {12581--12603},
  publisher = {Association for Computational Linguistics},
  doi = {10.18653/v1/2026.acl-long.573},
  note = {\url{https://aclanthology.org/2026.acl-long.573/}}
}

@misc{wang2026agentwm,
  author = {Liwen Wang and Zongjie Li and Yuchong Xie and Shuai Wang and
    Dongdong She and Wei Wang and Juergen Rahmel},
  title = {On Protecting Agentic Systems' Intellectual Property via
    Watermarking},
  year = {2026},
  howpublished = {arXiv preprint arXiv:2602.08401},
  eprint = {2602.08401},
  archivePrefix = {arXiv},
  primaryClass = {cs.AI},
  note = {\url{https://arxiv.org/abs/2602.08401}}
}

@inproceedings{meng2026acthook,
  author = {Wenlong Meng and Chen Gong and Terry Yue Zhuo and Fan Zhang and
    Kecen Li and Zheng Liu and Zhou Yang and Chengkun Wei and Wenzhi Chen},
  title = {Watermarking {LLM} Agent Trajectories},
  booktitle = {Proceedings of the 43rd International Conference on Machine
    Learning},
  year = {2026},
  series = {Proceedings of Machine Learning Research},
  volume = {306},
  pages = {87835--87857},
  publisher = {PMLR},
  note = {\url{https://proceedings.mlr.press/v306/meng26b.html}}
}

@misc{an2026seqwm,
  author = {Hyeseon An and Shinwoo Park and Dongsu Kim and Yo-Sub Han},
  title = {Sequential Behavioral Watermarking for {LLM} Agents},
  year = {2026},
  howpublished = {arXiv preprint arXiv:2605.11036},
  eprint = {2605.11036},
  archivePrefix = {arXiv},
  primaryClass = {cs.CR},
  note = {\url{https://arxiv.org/abs/2605.11036}}
}

@misc{gao2026trace,
  author = {Zheng Gao and Xiaoyu Li and Xiaoyan Feng and Jiaojiao Jiang and
    Yang Song and Yulei Sui and Zhenchang Xing and Liming Zhu},
  title = {{TRACE}: A Two-Channel Robust Attribution Watermark via
    Complementary Embeddings for {LLM}-Agent Trajectories},
  year = {2026},
  howpublished = {arXiv preprint arXiv:2607.08400},
  eprint = {2607.08400},
  archivePrefix = {arXiv},
  primaryClass = {cs.CR},
  note = {\url{https://arxiv.org/abs/2607.08400}}
}

@inproceedings{liu2024agentbench,
  author = {Xiao Liu and Hao Yu and Hanchen Zhang and Yifan Xu and Xuanyu Lei
    and Hanyu Lai and Yu Gu and Hangliang Ding and Kaiwen Men and Kejuan Yang
    and Shudan Zhang and Xiang Deng and Aohan Zeng and Zhengxiao Du and
    Chenhui Zhang and Sheng Shen and Tianjun Zhang and Yu Su and Huan Sun and
    Minlie Huang and Yuxiao Dong and Jie Tang},
  title = {{AgentBench}: Evaluating {LLM}s as Agents},
  booktitle = {The Twelfth International Conference on Learning
    Representations (ICLR)},
  year = {2024},
  note = {\url{https://openreview.net/forum?id=zAdUB0aCTQ}}
}

@inproceedings{debenedetti2024agentdojo,
  author = {Edoardo Debenedetti and Jie Zhang and Mislav Balunovi{\'c} and
    Luca Beurer-Kellner and Marc Fischer and Florian Tram{\`e}r},
  title = {{AgentDojo}: A Dynamic Environment to Evaluate Prompt Injection
    Attacks and Defenses for {LLM} Agents},
  booktitle = {Advances in Neural Information Processing Systems},
  year = {2024},
  volume = {37},
  pages = {82895--82920},
  doi = {10.52202/079017-2636},
  note = {Datasets and Benchmarks Track.
    \url{https://doi.org/10.52202/079017-2636}}
}

@misc{minisweagent,
  author = {{SWE-agent Team}},
  title = {{mini-swe-agent}},
  year = {2026},
  howpublished = {Software},
  note = {Version 2.4. Accessed: Nov. 18, 2026},
  url = {https://github.com/SWE-agent/mini-swe-agent}
}

@misc{opencode_current,
  author = {{OpenCode Contributors}},
  title = {{OpenCode}},
  year = {2026},
  howpublished = {Software},
  note = {Version 1.19.4. Accessed: Nov. 18, 2026},
  url = {https://github.com/anomalyco/opencode/releases/tag/v1.19.4}
}

@misc{openhands_sdk,
  author = {{OpenHands Contributors}},
  title = {{OpenHands Software Agent SDK}},
  year = {2026},
  howpublished = {Software},
  note = {{SDK} and tools version 1.58.1. Accessed: Nov. 18, 2026},
  url = {https://github.com/OpenHands/software-agent-sdk/releases/tag/v1.58.1}
}

@misc{li2026trustedari,
  author = {Qi Li and Zhenhua Zou and Shuo Li and Mingwei Xu and Zhuotao Liu},
  title = {{TrustedARI}: Towards Trust-Native Agentic Routing Infrastructure
    for Agentic {AI}},
  year = {2026},
  howpublished = {arXiv preprint arXiv:2606.15822},
  eprint = {2606.15822},
  archivePrefix = {arXiv},
  primaryClass = {cs.AI},
  note = {\url{https://arxiv.org/abs/2606.15822}}
}

@inproceedings{kirchenbauer2023watermark,
  author = {John Kirchenbauer and Jonas Geiping and Yuxin Wen and Jonathan Katz
    and Ian Miers and Tom Goldstein},
  title = {A Watermark for Large Language Models},
  booktitle = {Proceedings of the 40th International Conference on Machine
    Learning},
  year = {2023},
  series = {Proceedings of Machine Learning Research},
  volume = {202},
  pages = {17061--17084},
  publisher = {PMLR},
  note = {\url{https://proceedings.mlr.press/v202/kirchenbauer23a.html}}
}

@article{dathathri2024scalable,
  author = {Sumanth Dathathri and Abigail See and Sumedh Ghaisas and Po-Sen
    Huang and Rob McAdam and Johannes Welbl and Vandana Bachani and Alex
    Kaskasoli and Robert Stanforth and Tatiana Matejovicova and Jamie Hayes and
    Nidhi Vyas and Majd Al Merey and Jonah Brown-Cohen and Rudy Bunel and Borja
    Balle and Taylan Cemgil and Zahra Ahmed and Kitty Stacpoole and Ilia
    Shumailov and Ciprian Baetu and Sven Gowal and Demis Hassabis and Pushmeet
    Kohli},
  title = {Scalable Watermarking for Identifying Large Language Model Outputs},
  journal = {Nature},
  year = {2024},
  volume = {634},
  pages = {818--823},
  doi = {10.1038/s41586-024-08025-4},
  note = {\url{https://doi.org/10.1038/s41586-024-08025-4}}
}

@inproceedings{zhang2024remark,
  author = {Ruisi Zhang and Shehzeen Samarah Hussain and Paarth Neekhara and
    Farinaz Koushanfar},
  title = {{REMARK-LLM}: A Robust and Efficient Watermarking Framework for
    Generative Large Language Models},
  booktitle = {33rd USENIX Security Symposium (USENIX Security 24)},
  year = {2024},
  pages = {1813--1830},
  publisher = {USENIX Association},
  note = {\url{https://www.usenix.org/conference/usenixsecurity24/presentation/zhang-ruisi}}
}

@inproceedings{cohen2025adaptive,
  author = {Aloni Cohen and Alexander Hoover and Gabe Schoenbach},
  title = {Watermarking Language Models for Many Adaptive Users},
  booktitle = {2025 IEEE Symposium on Security and Privacy (SP)},
  year = {2025},
  pages = {2583--2601},
  doi = {10.1109/SP61157.2025.00084},
  note = {\url{https://doi.org/10.1109/SP61157.2025.00084}}
}

@inproceedings{qu2025multibit,
  author = {Wenjie Qu and Wengrui Zheng and Tianyang Tao and Dong Yin and Yanze
    Jiang and Zhihua Tian and Wei Zou and Jinyuan Jia and Jiaheng Zhang},
  title = {Provably Robust Multi-bit Watermarking for {AI-generated} Text},
  booktitle = {34th USENIX Security Symposium (USENIX Security 25)},
  year = {2025},
  pages = {201--220},
  publisher = {USENIX Association},
  note = {\url{https://www.usenix.org/conference/usenixsecurity25/presentation/qu-watermarking}}
}

@inproceedings{zhang2026character,
  author = {Zhaoxi Zhang and Xiaomei Zhang and Yanjun Zhang and He Zhang and
    Shirui Pan and Bo Liu and Asif Gill and Leo Yu Zhang},
  title = {Character-Level Perturbations Disrupt {LLM} Watermarks},
  booktitle = {Network and Distributed System Security Symposium (NDSS)},
  year = {2026},
  doi = {10.14722/ndss.2026.230138},
  note = {\url{https://doi.org/10.14722/ndss.2026.230138}}
}

@misc{sweagent_repository,
  author = {{SWE-agent Team}},
  title = {{SWE-agent}: Official Repository and Project Guidance},
  howpublished = {Software repository},
  note = {Accessed: Nov. 18, 2026},
  url = {https://github.com/SWE-agent/SWE-agent}
}

\onecolumn
\appendices
\section{Provider-Side Manipulation Results}
\label{app:robustness}
Table~\ref{tab:robustness} reports the complete Top-1 accuracy and MRR
results for the experiments in Section~\ref{sec:robustness}.

A2 prepends an identity-withholding notice, applies model-name redaction,
and normalizes whitespace outside fenced code blocks. Tool interactions,
intermediate messages, repository changes, and fenced code remain unchanged.

\medskip
\LidarTableRobustnessInline

\end{document}